\documentclass[twocolumn,tighten,times]{aastex631}

\usepackage{amsfonts}
\usepackage{mathrsfs}
\usepackage{bm}
\usepackage{lipsum}
\usepackage{physics}
\usepackage{multirow}
\usepackage{rotating}
\usepackage{graphicx} 
\usepackage{soul}
\usepackage{subfigure}
\usepackage{color}

\def\sunm{M_{\odot}}

\newcommand{\feii}{\ifmmode {\rm Fe\ II} \else Fe~{\sc ii}\fi}
\newcommand{\heii}{\ifmmode {\rm He\ II} \else He~{\sc ii}\fi}
\newcommand{\hei}{\ifmmode {\rm He\ I} \else He~{\sc i}\fi}
\newcommand{\oiii}{\ifmmode {\rm~[O\ III]} \else [O~{\sc iii}]\fi}
\newcommand{\nii}{\ifmmode {\rm~[N\ II]} \else [N~{\sc ii}]\fi}
\newcommand{\mgii}{\ifmmode {\rm~Mg\ II} \else Mg~{\sc ii}\fi}
\newcommand{\ciii}{\ifmmode {\rm~C\ III]} \else C~{\sc iii}]\fi}
\newcommand{\civ}{\ifmmode {\rm~C\ IV} \else C~{\sc iv}\fi}
\newcommand{\hb}{\rm{H$\beta$}}
\newcommand{\ha}{\rm{H$\alpha$}}
\newcommand{\hg}{\rm{H$\gamma$}}
\newcommand{\fab}{$F_{\rm H\alpha} / F_{\rm H\beta}$}
\newcommand{\mbh}{\ifmmode {M_{\bullet}} \else $M_{\bullet}$\fi}
\newcommand{\dotm}{\ifmmode {\dot{\mathscr{M}}} \else $\dot{\mathscr{M}}$\fi}
\def\ergs{\rm erg\,s^{-1}}
\def\ergscm{\rm erg\,s^{-1}\,cm^{-2}}
\def\ergscma{\rm erg\,s^{-1}\,cm^{-2}\,\AA^{-1}}
\def\kms{\rm km\,s^{-1}}
\def\mathdotM{\dot{\mathscr{M}}}

\shorttitle{Reverberation Mapping of NGC 4151}
\shortauthors{Feng et al.}

\begin{document}

\title{\bf \large Velocity-resolved Reverberation Mapping of Changing-look Active Galactic Nucleus NGC 4151 during Outburst Stage. II. Four Season Observation Results}

\author[0000-0002-1530-2680]{Hai-Cheng Feng} 
\affiliation{Yunnan Observatories, Chinese Academy of Sciences, Kunming 650216, Yunnan, People's Republic of China}
\affiliation{Key Laboratory for the Structure and Evolution of Celestial Objects, Chinese Academy of Sciences, Kunming 650216, Yunnan, People's Republic of China}
\affiliation{Center for Astronomical Mega-Science, Chinese Academy of Sciences, 20A Datun Road, Chaoyang District, Beijing 100012, People's Republic of China}
\affiliation{Key Laboratory of Radio Astronomy and Technology, Chinese Academy of Sciences, 20A Datun Road, Chaoyang District, Beijing 100101, People's Republic of China}

\author[0000-0003-3823-3419]{Sha-Sha Li}
\affiliation{Yunnan Observatories, Chinese Academy of Sciences, Kunming 650216, Yunnan, People's Republic of China}
\affiliation{Key Laboratory for the Structure and Evolution of Celestial Objects, Chinese Academy of Sciences, Kunming 650216, Yunnan, People's Republic of China}
\affiliation{Center for Astronomical Mega-Science, Chinese Academy of Sciences, 20A Datun Road, Chaoyang District, Beijing 100012, People's Republic of China}
\affiliation{Key Laboratory of Radio Astronomy and Technology, Chinese Academy of Sciences, 20A Datun Road, Chaoyang District, Beijing 100101, People's Republic of China}

\author{J. M. Bai}
\affiliation{Yunnan Observatories, Chinese Academy of Sciences, Kunming 650216, Yunnan, People's Republic of China}
\affiliation{Key Laboratory for the Structure and Evolution of Celestial Objects, Chinese Academy of Sciences, Kunming 650216, Yunnan, People's Republic of China}
\affiliation{Center for Astronomical Mega-Science, Chinese Academy of Sciences, 20A Datun Road, Chaoyang District, Beijing 100012, People's Republic of China}
\affiliation{Key Laboratory of Radio Astronomy and Technology, Chinese Academy of Sciences, 20A Datun Road, Chaoyang District, Beijing 100101, People's Republic of China}

\author[0000-0002-2153-3688]{H. T. Liu}
\affiliation{Yunnan Observatories, Chinese Academy of Sciences, Kunming 650216, Yunnan, People's Republic of China}
\affiliation{Key Laboratory for the Structure and Evolution of Celestial Objects, Chinese Academy of Sciences, Kunming 650216, Yunnan, People's Republic of China}
\affiliation{Center for Astronomical Mega-Science, Chinese Academy of Sciences, 20A Datun Road, Chaoyang District, Beijing 100012, People's Republic of China}

\author[0000-0002-2310-0982]{Kai-Xing Lu}
\affiliation{Yunnan Observatories, Chinese Academy of Sciences, Kunming 650216, Yunnan, People's Republic of China}
\affiliation{Key Laboratory for the Structure and Evolution of Celestial Objects, Chinese Academy of Sciences, Kunming 650216, Yunnan, People's Republic of China}
\affiliation{Center for Astronomical Mega-Science, Chinese Academy of Sciences, 20A Datun Road, Chaoyang District, Beijing 100012, People's Republic of China}

\author[0009-0005-3823-9302]{Yu-Xuan Pang}
\affiliation{Department of Astronomy, School of Physic, Peking University, Beijing 100871, People's Republic of China}
\affiliation{Kavli Institute for Astronomy and Astrophysics, Peking University, Beijing 100871, People's Republic of China}

\author[0000-0002-0771-2153]{Mouyuan Sun}
\affiliation{Department of Astronomy, Xiamen University, Xiamen, Fujian 361005, People’s Republic of China}

\author[0000-0003-4156-3793]{Jian-Guo Wang}
\affiliation{Yunnan Observatories, Chinese Academy of Sciences, Kunming 650216, Yunnan, People's Republic of China}
\affiliation{Key Laboratory for the Structure and Evolution of Celestial Objects, Chinese Academy of Sciences, Kunming 650216, Yunnan, People's Republic of China}
\affiliation{Center for Astronomical Mega-Science, Chinese Academy of Sciences, 20A Datun Road, Chaoyang District, Beijing 100012, People's Republic of China}

\author[0009-0000-7791-8192]{Yang-Wei Zhang}
\affiliation{South-Western Institute for Astronomy Research, Yunnan University, Kunming 650500, People’s Republic of China}

\author[0009-0005-2801-6594]{Shuying Zhou}
\affiliation{Department of Astronomy, Xiamen University, Xiamen, Fujian 361005, People’s Republic of China}

\correspondingauthor{Hai-Cheng Feng}
\email{hcfeng@ynao.ac.cn}

\begin{abstract}
We present the results of a four-year velocity-resolved reverberation mapping (RM) campaign of the changing-look active galactic nucleus (CL-AGN) NGC 4151 during its outburst phase. By measuring the time lags of the \ha, \hb, \hg, \hei, and \heii\ emission lines, we confirm a stratified broad-line region (BLR) structure that aligns with predictions from photoionization models. Intriguingly, we observed an ``anti-breathing" phenomenon, where the lags of broad emission lines decreased with increasing luminosity, contrary to the typical expectation. This anomaly may be attributed to the influence of the ultraviolet-optical lag or non-virialized motions in the BLR gas. Velocity-resolved RM and ionization mapping analyses revealed rapid and significant changes in the BLR geometry and kinematics on timescales within one year, which cannot be interpreted by any single mechanism, such as an inhomogeneous BLR, variations in radiation pressure, or changes in the illuminated ionizing field. Additionally, the \hb\ lags of NGC 4151 and other CL-AGNs agree with the radius-luminosity relationship established for AGNs with low accretion rates, implying that the CL phenomenon is more likely driven by intrinsic changes in the accretion rate rather than obscuration. These findings provide new insights into the complex internal processes of CL-AGNs and highlight the importance of long-term, multi-line RM for understanding BLR structures, geometry, and kinematics.

\end{abstract}


\section{Introduction} \label{sec:1}
Active galactic nuclei (AGNs) are the compact and luminous regions at the centers of galaxies, powered by the accretion process of supermassive black holes \citep[SMBHs;][]{Lynden-Bell1969, Rees1984}. Measuring the mass of SMBHs and studying their surrounding environments are crucial for understanding a series of astrophysical phenomena, such as black hole accretion physics \citep{Shakura1973}, the formation of relativistic jets \citep{Blandford1977}, and the co-evolution of SMBHs and their host galaxies \citep{Kormendy2013}.

Over the past few decades, extensive observational and theoretical studies have revealed the basic picture of AGN structure: the central SMBH produces most of the observed continuum radiation through an accretion disk, which is surrounded by the broad-line region (BLR) composed of rapidly orbiting ionized gas emitting broad emission lines, a dust torus that can obscure the inner regions depending on the viewing angle, and the narrow-line region (NLR) extending to larger scales and emitting narrow emission lines \citep{Antonucci1993, Urry1995}. Despite the huge success of the unified model in explaining the diverse observational features of AGNs, there remain many mysteries regarding the detailed physical processes and evolution of these structures, such as the observed accretion disk sizes often being larger than theoretically expected, the complex geometry and kinematics of the BLR, and the influence of radiation pressure on the internal physical properties of AGNs \citep[e.g.,][]{Gaskell2018, Pancoast2018, Villafana2022, Chen2023, Liu2024, Li2024, Zhou2024}.

Variability provides an important avenue for exploring the inner structure of AGNs. By analyzing the variability characteristics of AGNs across different wavelengths, it is possible to effectively probe the sizes, structures, and dynamical properties of distinct physical regions within the AGN. In particular, the technique of reverberation mapping \citep[RM;][]{Blandford1982, Peterson1993} utilizes the time delay ($\tau$) between the emission lines and the continuum variability, providing a powerful tool for measuring the location of the BLR. This method is based on the photoionization model assumption: the broad emission lines are produced by the photoionization of BLR gas by the central continuum radiation, hence the time delay in emission lines relative to the continuum reflects the travel time of ionizing photons from the central ionizing source to the BLR. By deriving the characteristic velocity ($v$) of the BLR gas from the broadening of the emission lines, we can further estimate the SMBH mass using the formula \citep{Peterson2004}:
\begin{equation}\label{equ1}
\mbh = \frac{c \tau v^2}{G},
\end{equation}
where $c$ is the speed of light, and $G$ is the gravitational constant.

To date, the BLR size and SMBH mass for more than 100 AGNs have been measured using the RM method \citep{Peterson2004, Bentz2013, Oknyansky2021, Villafana2022, Cho2023, Nagoshi2024, Zastrocky2024}. Importantly, these measurements have revealed a tight correlation between the AGN luminosity ($L$) and the BLR radius ($R$) \citep{Kaspi2000, Bentz2013}:
\begin{equation}
R \propto L^{\alpha},
\end{equation}
where the exponent $\alpha$ ranges between 0.5 and 0.7. Leveraging this relationship (hereafter the $R-L$ relation), the mass of the SMBH can be inferred from a single-epoch spectroscopic observation, making it widely applicable in AGN research.

Additionally, some high-quality RM data allow for a detailed analysis of the emission line response at different line-of-sight velocities, enabling the inference of the geometry and kinematic properties of the BLR. This approach is known as the ``velocity-resolved" RM method. These studies have revealed that, in addition to Keplerian motion, the BLR often exhibits complex kinematic features such as inflows or outflows \citep{Denney2009, Bentz2010, Grier2013, Pancoast2014, Du2016, Lu2019, Li2021}. Intriguingly, RM observations of the same AGN at different durations sometimes show significant changes in the kinematic features of the BLR \citep{Grier2017, Lu2022, Chen2023, Yao2024}. It is generally believed that changes in BLR kinematics may be related to variations in the accretion rate \citep{Proga2000, Kollatschny2013}, where higher accretion rates could lead to stronger radiation pressure driving more outflowing gas. However, \citet{Du2016} found no significant correlation between BLR kinematics and accretion rates through a sample of high accretion rate AGNs. This phenomenon might be due to the substantial differences in the ionization states and gas distributions of the BLR among different AGNs, making it challenging to clearly reflect the effects of radiation pressure or accretion rate in the existing RM samples. On the other hand, the velocity-resolved ionization mapping (IM) technique proposed by \citet{Li2024} suggests that some observed velocity-resolved lags cannot be fully explained by simple inflow or outflow motions within a symmetric BLR, indicating a need for more complex geometric or kinematic configurations.

To gain a deeper understanding of the impact of radiation pressure from the central continuum on the BLR gas, we need to quantitatively study the inherent connection between its kinematic properties and accretion rates. An effective approach is to conduct long-term, multi-epoch, and multi-line velocity-resolved RM monitoring of the same AGN under different luminosity states. By systematically analyzing how the spatial distribution and velocity field of the BLR gas respond to changes in luminosity, we may be able to elucidate the role of radiation pressure in regulating the physical properties of the BLR. Currently, such long-term high-quality monitoring data are relatively scarce. The few AGNs with multiple observations suffer from the following issues: 1) long intervals between observations, making it impossible to track the evolution of BLR properties during these periods \citep{Denney2009, Feng2021a}; 2) some targets with several years of continuous observations often have limited variations in luminosity, making it challenging to quantitatively assess the impact of radiation pressure \citep{Lu2022}. This results in a lack of a comprehensive and clear understanding of the temporal evolution of the physical properties of the BLR.

\begin{figure*}[!ht]
\centering 
\includegraphics[scale=0.65]{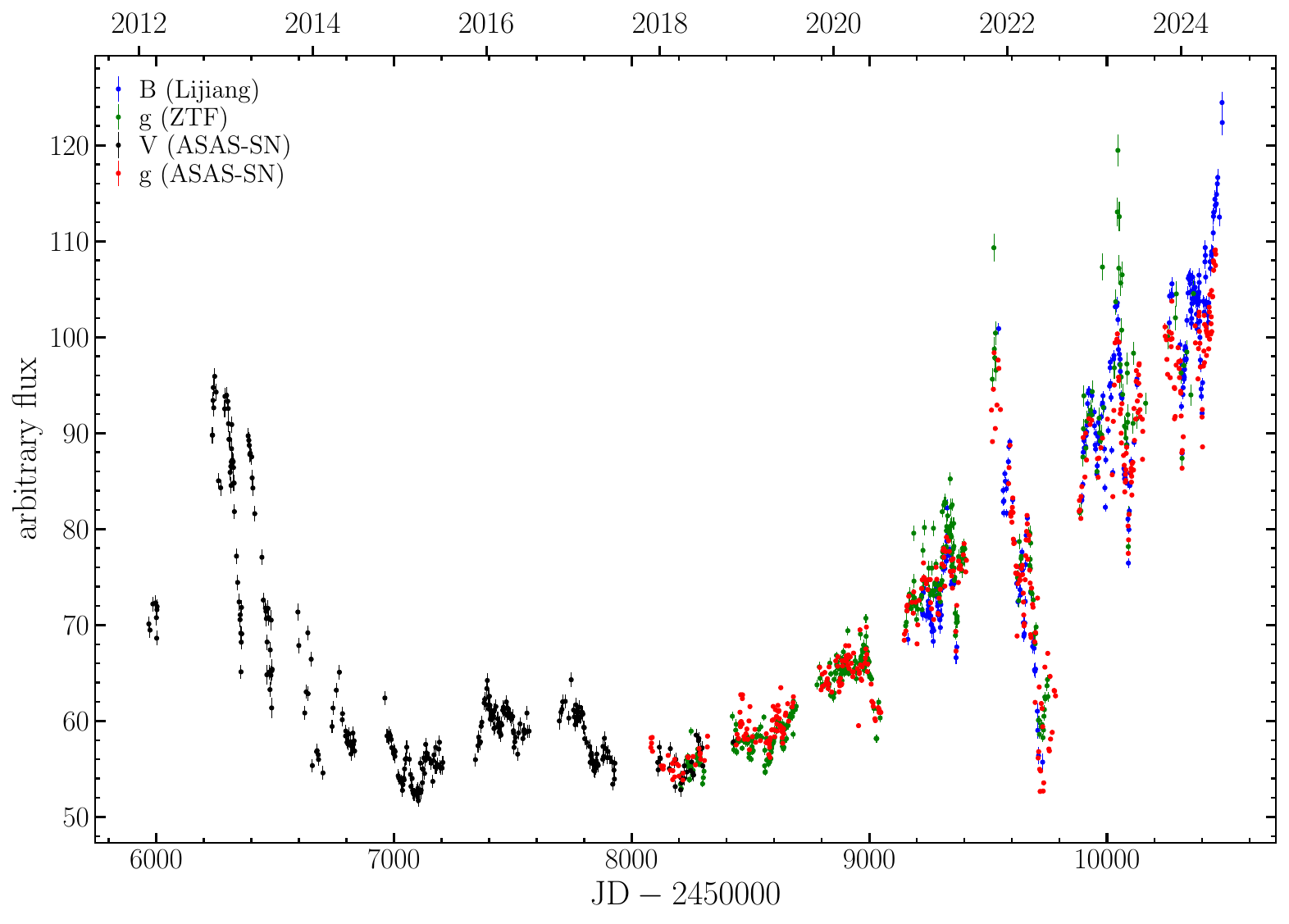}
\caption{Long-term light curve of NGC 4151. Data from different bands were intercalibrated using PyCALI, with ASAS-SN $g$-band data as the reference.
}
\label{fig1}
\end{figure*}

NGC 4151 is a nearby ($z$ = 0.003326) Seyfert galaxy that has historically undergone several significant luminosity variations, and often accompanied by changes in spectral type, i.e., the Changing-Look \citep[CL;][]{Wang2019, Guo2024, Wang2024} phenomenon. These characteristics suggest that the kinematic properties of its BLR may also undergo changes, thereby altering the distribution of gas. Therefore, we conducted an RM observation of NGC 4151 in 2020-2021, which revealed that its velocity-resolved lags were indeed inconsistent with previous results \citep[see][hereafter Paper I]{Li2022}. Subsequently, \citet{Chen2023} further confirmed the long-term kinematic changes in the BLR of NGC 4151 by collecting archival data and combining them with their own observations. Recently, this target has re-entered an active phase, brightening continuously for over 6.5 years (see Figure~\ref{fig1}), with each observing season exhibiting sufficient variability to allow for the detection of RM signals. This is an excellent opportunity to study the internal physical properties of AGNs in depth, motivating us to initiate a 4-year spectroscopic and multi-band photometric observing campaign. The primary goals of this project are to analyze the timescales of changes in the geometry and kinematics of the BLR through continuous monitoring, assess the applicability of velocity-resolved RM, explore the physical properties of the accretion disk, understand the mechanisms behind CL-AGNs, and investigate whether these changes are related to the accretion rate.

In this paper, we present the basic results of recent RM measurements for multiple broad emission lines over the past four observing seasons, along with the related observational properties in CL-AGNs. In Section \ref{sec:2}, we outline the observation and data processing procedures. In Section \ref{sec:3}, we describe the analysis and results of these data. In Section \ref{sec:4}, we provide a detailed discussion of our findings. Finally, in Section \ref{sec:5}, we summarize the main conclusions of the paper. Throughout this paper, we adopt a cosmology with $H_0 = 67$ km s$^{-1}$ Mpc$^{-1}$, $\Omega_m = 0.32$, and $\Omega_\Lambda=0.68$ \citep{PlanckCollaboration2020}.

\begin{figure*}[!ht]
\centering 
\includegraphics[scale=0.65]{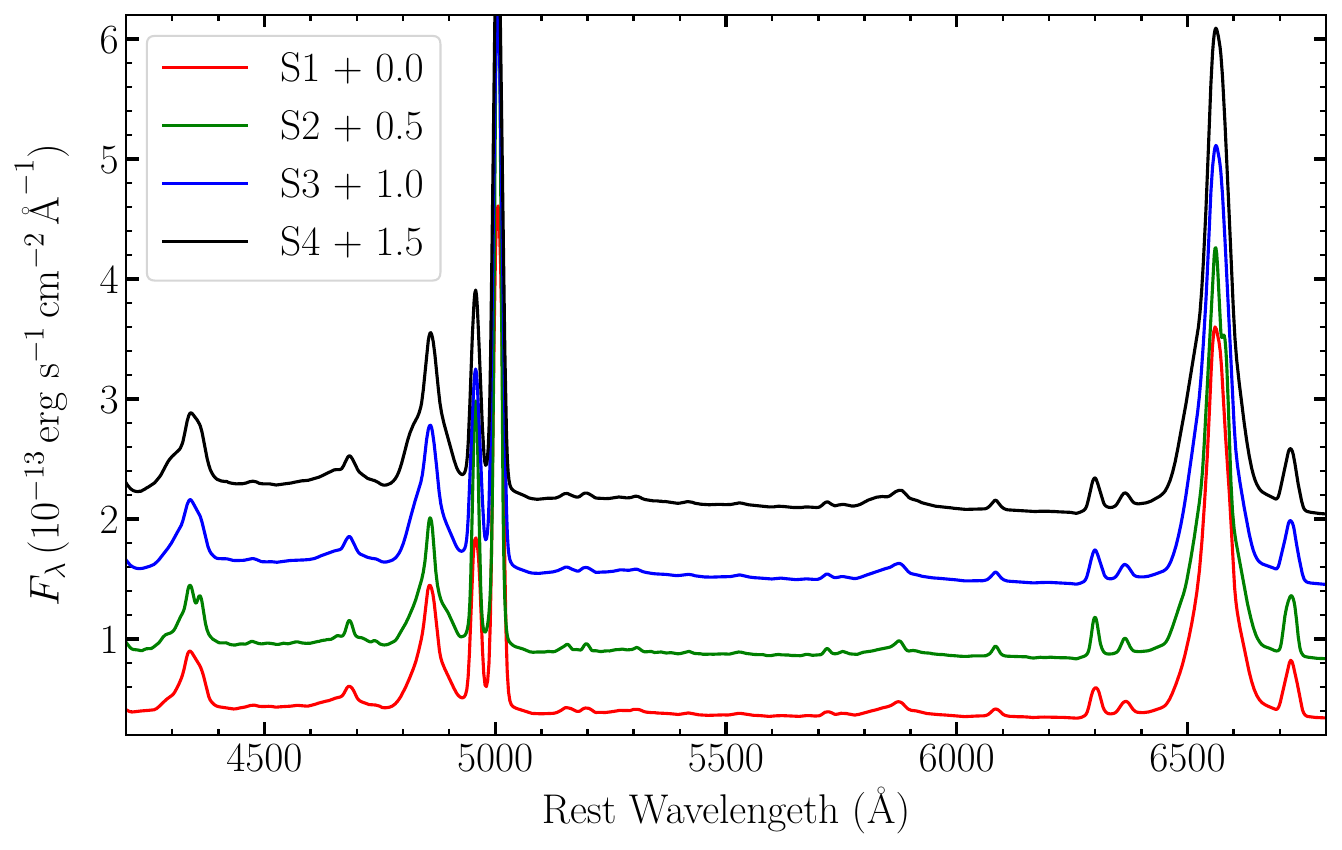}
\caption{Mean spectrum for each observing season. The spectra are vertically shifted by an interval of $0.5 \times 10^{-13} \ergscma$ to facilitate distinction, with red, green, blue, and black representing S1, S2, S3, and S4, respectively.
}
\label{fig2}
\end{figure*}

\section{Data} \label{sec:2}
\subsection{Lijiang Observations and Data Reduction} \label{sec:2.1}
To ensure the homogeneity and comparability of data across multiple years, our observing strategy and data reduction procedures are nearly consistent with that of Paper I, with adjustments made to the photometric filters and the combinations of grisms and slits according to the specific observational purposes. Here, we briefly summarize our observation scheme and describe the differences for each observing season.

\begin{deluxetable}{lcccccccccc}[!ht]
\tablecolumns{10}
\tablewidth{\textwidth}
\tabletypesize{\scriptsize}
	\tablecaption{Details of Observations for Each Observing Season \label{tab:1}}	
	\tablewidth{\textwidth}
		\tablehead{\colhead{Season} &
			\colhead{Duration} &
			\colhead{Filter(s)} &			
			\colhead{Grism} &
			\colhead{Slit Width}
		}
\startdata
S1 & 2020.11.08 -- 2021.06.02 & $B$  & Grism 3 & 2\farcs5 \\
S2 & 2021.11.25 -- 2022.05.29 & $B$  & Grism 14 & 2\farcs5 \\
S3 & 2022.11.10 -- 2023.07.01 & $B$, $R$  & Grism 14 & 5\farcs05 \\
S4 & 2023.11.13 -- 2024.06.23 & $B$, $V$, $R$, $I$  & Grism 14 & 5\farcs05 \\
\enddata
\end{deluxetable}

The observations presented in this paper were conducted from 2020 to 2024 using the Lijiang 2.4 m optical telescope at the Yunnan Observatories of the Chinese Academy of Sciences. Each observing season for NGC 4151 approximately spans from November to May of the following year. All observations were carried out using the Yunnan Faint Object Spectrograph and Camera, which can rapidly switch between photometric and spectroscopic modes. During all four observing seasons, we simultaneously conducted photometric and spectroscopic monitoring. The photometric observations employed Cousins-Johnson filters, while the spectroscopic observations used long-slit spectroscopy. For each spectroscopic observation, we rotated the slit to include a comparison star near the target source in the slit, ensuring accurate flux calibration. Additionally, we used a UV-blocking filter to eliminate second-order spectra affecting \ha.

In the first observing season (S1), we used the $B$ band for photometric monitoring, primarily to obtain high-quality continuum light curves. Spectroscopic observations employed a 2\farcs5 slit to minimize the impact of seeing while maximizing resolution. We used Grism 3 with a dispersion of 2.96 \AA\,pixel$^{-1}$ to achieve a broad wavelength coverage (3800-9000 \AA) and high throughput. In the second observing season (S2), to achieve higher spectral resolution and obtain more detailed BLR geometry and kinematics information, we used Grism 14 with a dispersion of 1.74 \AA\,pixel$^{-1}$. In the third observing season (S3), aiming to further measure the accretion disk size, which requires high flux accuracy, we used a wider 5\farcs05 slit and added the $R$ band for photometric monitoring. In the fourth observing season (S4), we further included $V$ and $I$ bands in the photometric monitoring to better constrain the accretion disk size measurements. Detailed observational information for each year is listed in Table \ref{tab:1}.

Data reduction was performed using the PyRAF software \citep{Pyraf2012}, including bias subtraction, flat-fielding, wavelength calibration, spectral and photometric flux extraction, and flux calibration using spectrophotometric standard stars. The extraction aperture radius for both photometric and spectroscopic data was 4\farcs24. For photometric data, we used several comparison stars near the target for relative flux calibration. For spectroscopic data, we obtained the fiducial spectrum of the comparison star during good weather conditions and used it for absolute flux calibration. Figure \ref{fig2} presents the mean spectrum for each observing season. Note that our continuum measurements only utilize the $B$-band data, while other photometric data will be used in future analyses of the accretion disk size.

\subsection{Archival Time-domain Photometric Data} \label{sec:2.2}
We also utilized publicly available photometric survey data from the All-Sky Automated Survey for Supernovae\footnote{\url{https://www.astronomy.ohio-state.edu/asassn/index.shtml}} \citep[ASAS-SN;][]{Shappee2014, Hart2023} and the Zwicky Transient Facility\footnote{\url{https://www.ztf.caltech.edu/}} \citep[ZTF;][]{Masci2019} to investigate the long-term variability of NGC 4151 over the past decade. Both projects offer high sampling cadence and temporal overlap with our RM observations, enabling us to extend the time baseline of the continuum light curves and enhance the sampling cadence for measuring time lags.

\begin{deluxetable*}{lcccccccccc}[!htbp]
 \tablecolumns{10}
\tablewidth{\textwidth}
\tabletypesize{\scriptsize}
\tablecaption{Light Curves of Emission Lines and Photometric Continuum}
\label{tab:2}
\tablehead{\multicolumn{6}{c}{Spectra}        &
      \colhead{}                         &
      \multicolumn{3}{c}{Photometry}     \\ 
      \cline{1-6} \cline{8-10}  
      \colhead{JD - 2,459,000}                       &
      \colhead{$F_{\rm H\alpha}$}               &
      \colhead{$F_{\rm H\beta}$}         &
      \colhead{$F_{\rm H\gamma}$}               &
      \colhead{$F_\hei$}               &
      \colhead{$F_\heii$}               &
      \colhead{}                         &
      \colhead{JD - 2,459,000}     &
      \colhead{mag}          &
      \colhead{$\rm Obs$}             
      }
\startdata
\multicolumn{10}{@{}c@{}}{S1} \\
\hline
162.44 & $147.136 \pm 1.821$ & $46.207 \pm 0.725$ & $19.227 \pm 0.626$ & $7.533 \pm 0.216$ & $8.290 \pm 1.395$ & & 153.05 & $12.959 \pm 0.020$ & ZTF \\
224.38 & $163.628 \pm 1.828$ & $51.484 \pm 0.735$ & $22.135 \pm 0.652$ & $8.387 \pm 0.244$ & $13.503 \pm 1.400$ & & 156.02 & $12.959 \pm 0.020$ & ZTF \\
225.37 & $164.933 \pm 1.826$ & $52.237 \pm 0.732$ & $21.685 \pm 0.641$ & $8.742 \pm 0.236$ & $14.133 \pm 1.399$ & & 158.02 & $12.924 \pm 0.019$ & ZTF \\
230.31 & $167.195 \pm 1.830$ & $53.336 \pm 0.738$ & $23.489 \pm 0.661$ & $8.966 \pm 0.255$ & $14.731 \pm 1.402$ & & 162.43 & $12.903 \pm 0.017$ & Lijiang \\
239.30 & $169.726 \pm 1.833$ & $52.926 \pm 0.742$ & $22.769 \pm 0.671$ & $9.259 \pm 0.265$ & $13.153 \pm 1.404$ & & 166.01 & $12.867 \pm 0.019$ & ZTF \\
\hline
\multicolumn{10}{@{}c@{}}{S2} \\
\hline
544.44 & $206.077 \pm 2.541$ & $65.104 \pm 0.722$ & $33.918 \pm 0.875$ & $9.628 \pm 0.356$ & $31.878 \pm 1.311$ & & 514.14 & $12.356 \pm 0.015$ & ASAS-SN \\
546.42 & $201.967 \pm 2.541$ & $66.329 \pm 0.720$ & $32.534 \pm 0.871$ & $10.126 \pm 0.354$ & $28.044 \pm 1.307$ & & 518.13 & $12.420 \pm 0.013$ & ASAS-SN \\
560.46 & $201.670 \pm 2.541$ & $64.198 \pm 0.721$ & $31.154 \pm 0.872$ & $9.948 \pm 0.355$ & $20.261 \pm 1.307$ & & 522.12 & $12.315 \pm 0.016$ & ASAS-SN \\
565.38 & $198.503 \pm 2.540$ & $61.997 \pm 0.719$ & $27.091 \pm 0.868$ & $9.407 \pm 0.352$ & $13.374 \pm 1.305$ & & 525.11 & $12.249 \pm 0.017$ & ASAS-SN \\
566.43 & $199.295 \pm 2.540$ & $62.398 \pm 0.719$ & $28.255 \pm 0.868$ & $9.545 \pm 0.352$ & $13.816 \pm 1.305$ & & 530.11 & $12.393 \pm 0.014$ & ASAS-SN \\
\hline
\multicolumn{10}{@{}c@{}}{S3} \\
\hline
895.43 & $195.206 \pm 1.705$ & $63.106 \pm 0.453$ & $25.636 \pm 0.639$ & $8.840 \pm 0.302$ & $14.609 \pm 1.473$ & & 883.01 & $12.580 \pm 0.010$ & ASAS-SN \\
898.44 & $195.387 \pm 1.705$ & $64.458 \pm 0.453$ & $26.805 \pm 0.640$ & $8.906 \pm 0.303$ & $14.697 \pm 1.473$ & & 884.15 & $12.553 \pm 0.010$ & ASAS-SN \\
903.42 & $195.189 \pm 1.705$ & $64.406 \pm 0.452$ & $28.377 \pm 0.639$ & $9.583 \pm 0.302$ & $19.712 \pm 1.474$ & & 886.15 & $12.577 \pm 0.009$ & ASAS-SN \\
908.42 & $197.442 \pm 1.705$ & $65.276 \pm 0.451$ & $28.484 \pm 0.637$ & $10.268 \pm 0.301$ & $19.267 \pm 1.473$ & & 890.14 & $12.599 \pm 0.010$ & ASAS-SN \\
910.44 & $197.459 \pm 1.705$ & $65.478 \pm 0.451$ & $28.771 \pm 0.637$ & $10.451 \pm 0.301$ & $16.693 \pm 1.472$ & & 891.13 & $12.543 \pm 0.008$ & ASAS-SN \\
\hline
\multicolumn{10}{@{}c@{}}{S4} \\
\hline
1263.43 & $213.015 \pm 2.190$ & $78.809 \pm 1.042$ & $31.368 \pm 1.784$ & $12.522 \pm 0.440$ & $29.009 \pm 2.627$ & & 1244.15 & $12.192 \pm 0.007$ & ASAS-SN \\
1271.44 & $226.202 \pm 2.190$ & $81.539 \pm 1.042$ & $34.311 \pm 1.783$ & $14.105 \pm 0.439$ & $26.484 \pm 2.627$ & & 1245.14 & $12.207 \pm 0.005$ & ASAS-SN \\
1273.42 & $228.691 \pm 2.190$ & $82.495 \pm 1.043$ & $34.395 \pm 1.785$ & $14.529 \pm 0.440$ & $28.231 \pm 2.627$ & & 1250.13 & $12.213 \pm 0.005$ & ASAS-SN \\
1275.41 & $225.322 \pm 2.190$ & $83.014 \pm 1.043$ & $37.298 \pm 1.787$ & $14.961 \pm 0.440$ & $30.960 \pm 2.628$ & & 1252.14 & $12.247 \pm 0.005$ & ASAS-SN \\
1307.34 & $228.303 \pm 2.191$ & $81.345 \pm 1.045$ & $38.135 \pm 1.792$ & $14.343 \pm 0.443$ & $30.156 \pm 2.630$ & & 1255.12 & $12.274 \pm 0.005$ & ASAS-SN \\
\enddata
\tablecomments{The emission-line flux is in units of $10^{-13}~\ergscm$. The photometric data include $B$-band observations from Lijiang and the $g$-band observations from ZTF and ASAS-SN. These data have been intercalibrated to the $B$-band using PyCALI and are labeled with the corresponding telescope in the `Obs' column.
\\
(This table is available in a machine-readable form in the online journal.)}
\end{deluxetable*}

ASAS-SN is a wide-field monitoring project consisting of 24 globally distributed telescopes, capable of imaging the entire sky nightly. The project first observed NGC 4151 on 2012 February 11 and provided continuous $V$-band light curves until 2018 November 26. From 2017 November 23 to the present, observations were conducted in the $g$-band. In this work, we utilized all the available data up to 2024 May 27.

ZTF employs a 48-inch Schmidt telescope to scan approximately 3760 square degrees of the northern hemisphere in the $g$, $r$, and $i$ bands, achieving a limiting magnitude exceeding 20 and a cadence of about two days. We obtained the light curves of the object from the latest Data Release 21, covering observations from 2018 March 25 to 2024 February 23.

Since our $B$-band wavelength coverage is closer to the $g$-band of ASAS-SN and ZTF, we focused our analysis on the $g$-band data while using the ASAS-SN $V$-band to supplement data before 2017. For the long-term variability analysis, we primarily relied on the ASAS-SN $g$-band data as a reference. RM measurements for the first year were directly obtained from Paper I, with ZTF $g$-band light curves retrieved via Automatic Learning for the Rapid Classification of Events (ALeRCE) \footnote{\url{https://alerce.online/}} \citep{Forster2021, Sanchez-Saez2021}. In the following three observing seasons, due to the lower sampling rate of the ZTF data, we mainly used the ASAS-SN $g$-band observations. Before analysis, data points with signal-to-noise ratio (S/N) below 5 were excluded, and observations were binned within 0.8-day intervals for each dataset.


\begin{figure*}
\centering 
\includegraphics[scale=0.93]{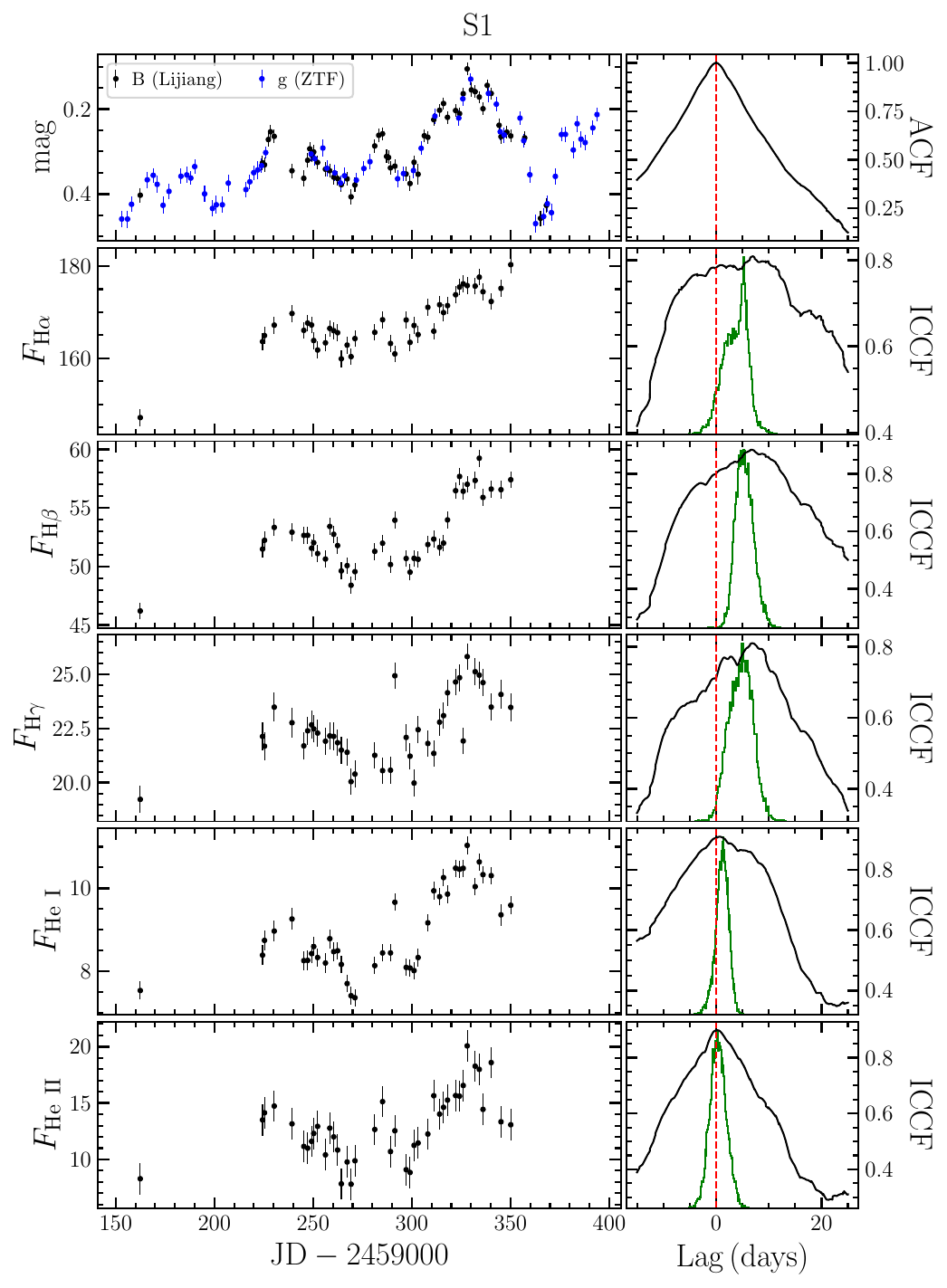}
\caption{Light curves and ICCF analysis results for each observing season. In each ICCF panel, the black line is the ICCF, the green histogram represents the CCCD, and the red dashed line denotes the zero lag.
}
\label{fig3}
\end{figure*}

\begin{figure*}
\centering 
  \figurenum{3}
\includegraphics[scale=0.93]{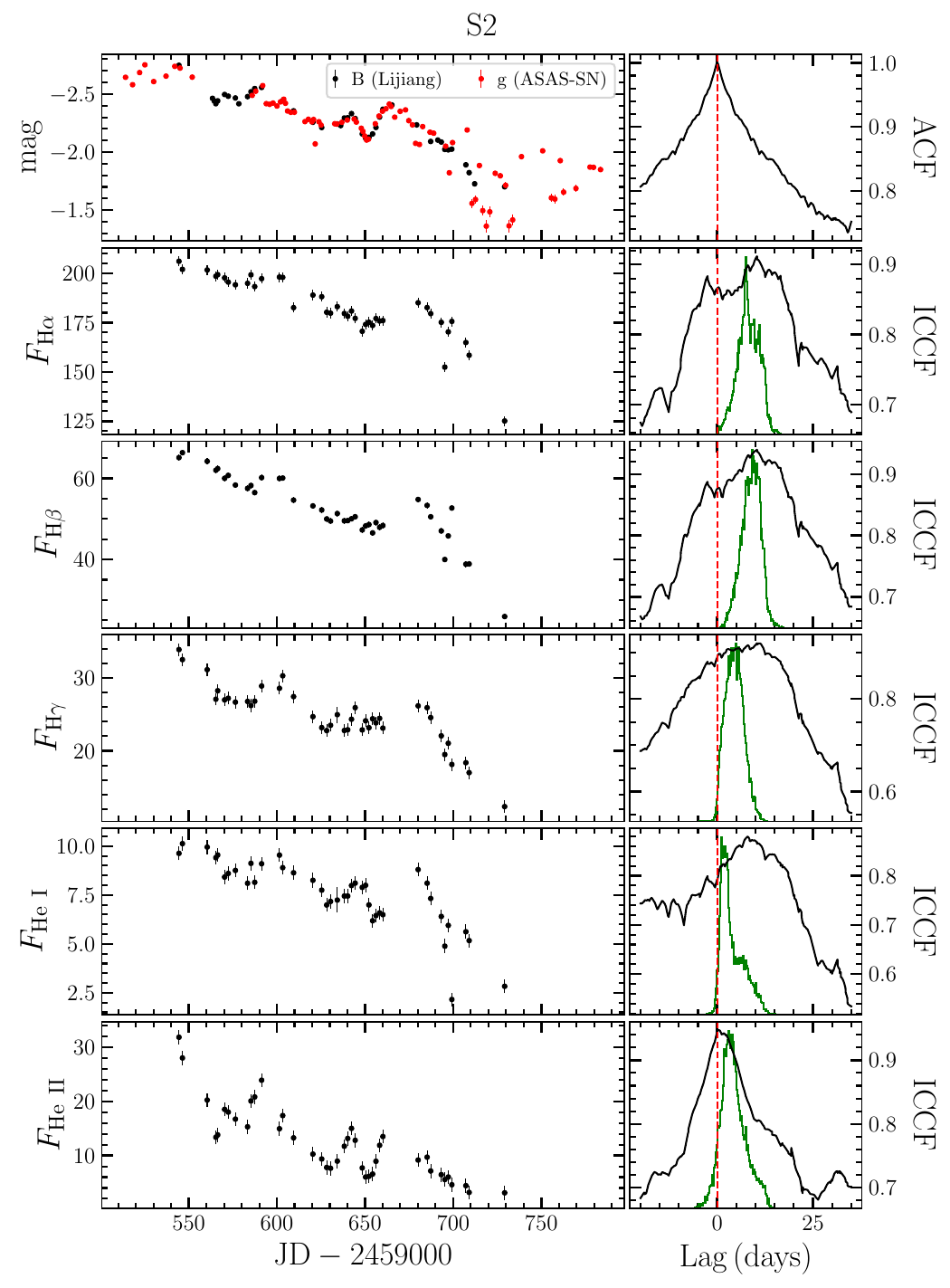}
\caption{(continued).
}
\end{figure*}

\begin{figure*}
\centering 
  \figurenum{3}
\includegraphics[scale=0.93]{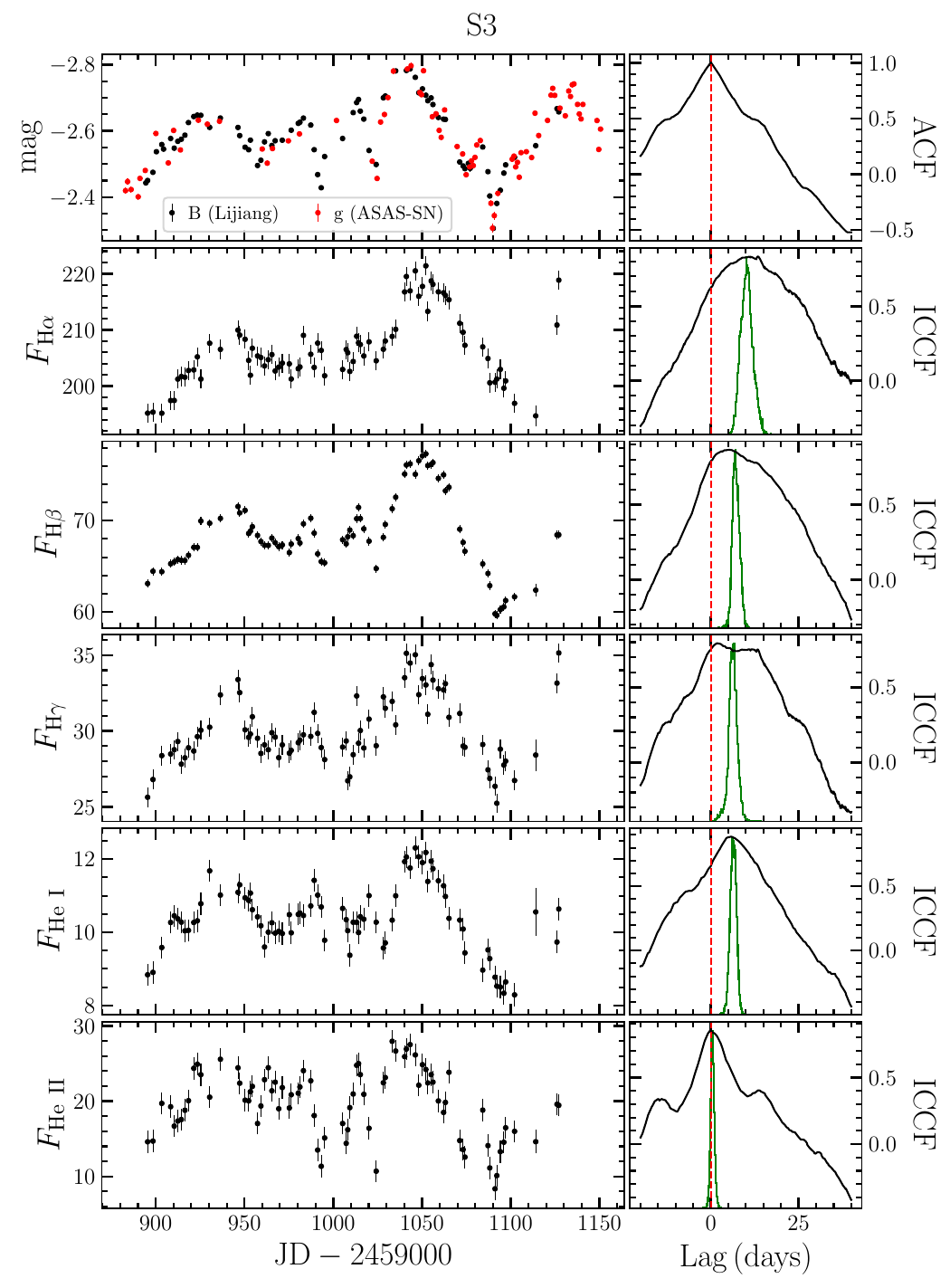}
\caption{(continued).
}
\end{figure*}

\begin{figure*}
\centering 
  \figurenum{3}
\includegraphics[scale=0.93]{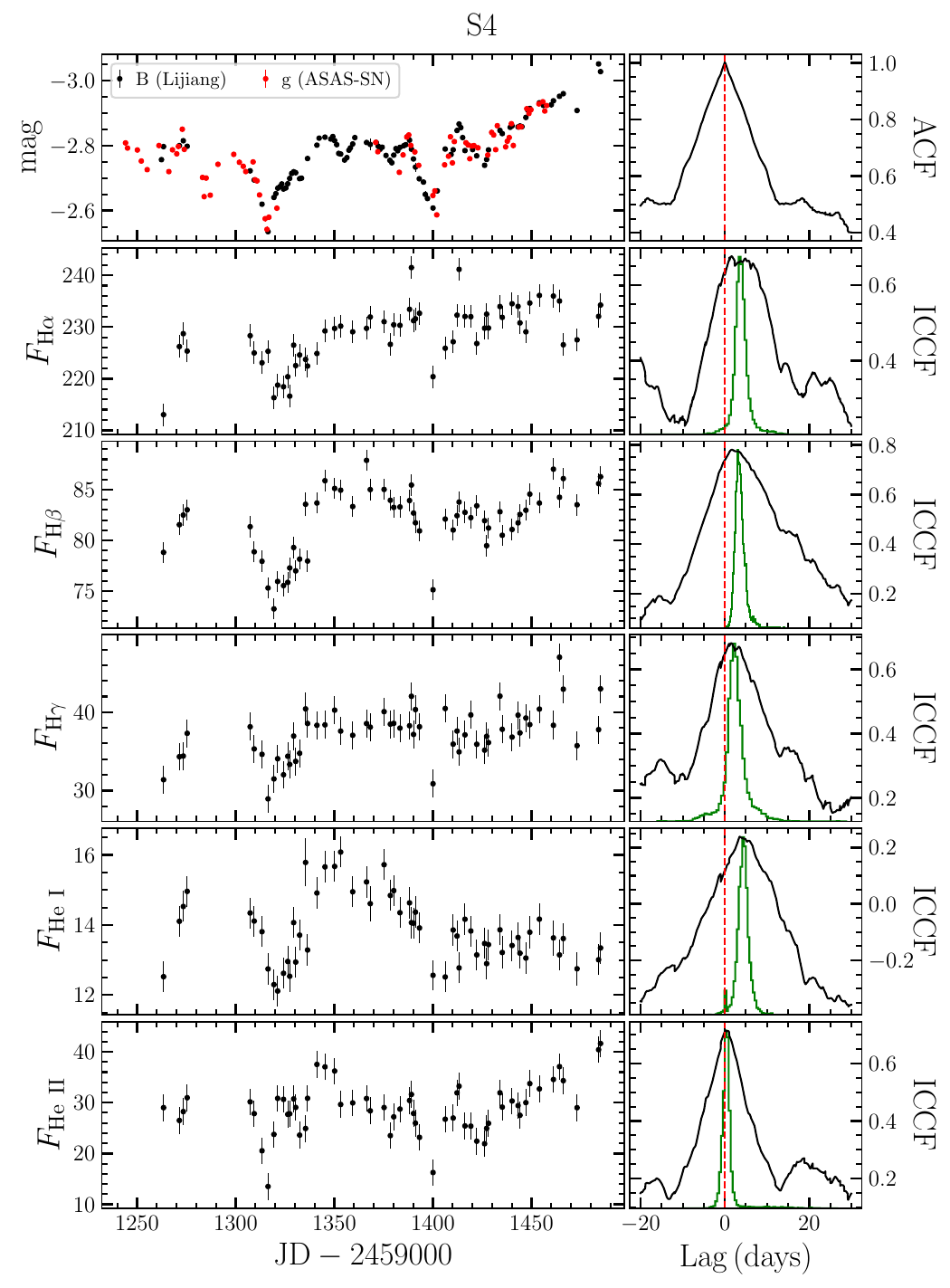}
\caption{(continued).
}
\end{figure*}

\section{Analysis and Results} \label{sec:3}
\subsection{Intercalibration of Photometric Data} \label{sec:3.1}
Our long-term light curve and RM continuum are obtained from multiple photometric projects. Differences in filters/telescopes/CCD transmission and data processing methods across these projects can lead to systematic offsets in the contribution of the AGN and host galaxy. Therefore, it is necessary to intercalibrate all data to a uniform light curve before analysis.

\begin{deluxetable*}{lcccccccccc}[!ht]
 \tablecolumns{10}
\tablewidth{\textwidth}
\tabletypesize{\scriptsize}
	\tablecaption{Properties of RM-Observed CL-AGNs \label{tab:3}}	
	\tablewidth{\textwidth}
		\tablehead{\colhead{Object} &
			\colhead{$\tau_{\rm H\beta}$} &
			\colhead{${\rm log}\, L_{\rm 5100}$} &			
			\colhead{${\rm log} ~(\mbh/M_{\odot})$} &
			\colhead{$\rm log~\mathdotM$} &
   \colhead{References} \\
   &\multicolumn{1}{c}{(days)}  &\multicolumn{1}{c}{($\rm erg~s^{-1}$)} 
		}
\startdata
NGC 4151 (S1) & $4.95_{-1.20}^{+2.12}$ &$42.66 \pm 0.09$  & $7.37_{-0.10}^{+0.19}$ &$-1.27_{-0.25}^{+0.39}$ & This work \\
NGC 4151 (S2)& $9.77_{-3.55}^{+1.40}$ & $42.60\pm0.18$ & $7.71_{-0.16}^{+0.06}$&$-2.01_{-0.41}^{+0.30}$& This work  \\
NGC 4151 (S3)& $7.32_{-1.12}^{+0.88}$ & $42.85 \pm 0.05$ & $7.56_{-0.07}^{+0.05}$ & $-1.36_{-0.15}^{+0.13}$& This work  \\
NGC 4151 (S4)& $3.25_{-0.72}^{+1.40}$ & $42.94 \pm 0.04$ & $7.27_{-0.10}^{+0.19}$ & $-0.63_{-0.20}^{+0.38}$ & This work \\
NGC 4151 & $6.59_{-0.76}^{+1.12}$ & $42.09 \pm 0.21$ & $7.72_{-0.06}^{+0.07}$ & $-2.81_{-0.57}^{+0.37}$& 1, 2, 3 \\
NGC 3516$^{a}$ & $7.50_{-0.77}^{+2.05}$ & $42.45 \pm  0.12$ & $7.38_{-0.05}^{+0.13}$&  $-1.60_{-0.21}^{+0.31}$ & 4\\
NGC 3516& $11.68_{-1.53}^{+1.02}$ & $42.79 \pm  0.20$ & $7.82_{-0.08}^{+0.05}$& $-1.97_{-0.52}^{+0.41}$ & 2, 3, 5\\
NGC 2617$^{a}$ & $5.38_{-1.12}^{+0.98}$  & $42.71 \pm 0.06$ & $7.32_{-0.09}^{+0.08}$ & $-1.09_{-0.20}^{+0.18}$ & 6\\
NGC 2617& $4.32_{-1.35}^{+1.10}$  & $42.67 \pm 0.16$ & $7.74_{-0.17}^{+0.11}$ & $-1.98_{-0.51}^{+0.55}$ & 3, 7\\ 
Fairall 9& $17.4_{-4.3}^{+3.2}$  & $43.98 \pm 0.04$ & $8.09_{-0.12}^{+0.07}$ & $-0.71_{-0.21}^{+0.31}$ & 2, 3, 8\\ 
Mrk 590& $20.7_{-2.7}^{+3.5}$  & $43.59 \pm 0.06$ & $7.50_{-0.06}^{+0.07}$ & $-0.22_{-0.25}^{+0.24}$ & 2, 3, 9\\ 
Mrk 590& $14.0_{-8.8}^{+8.5}$  & $43.14 \pm 0.09$ & $7.58_{-0.48}^{+0.22}$ & $-0.91_{-0.30}^{+0.28}$ & 2, 3, 9\\ 
Mrk 590& $29.2_{-5.0}^{+4.9}$  & $43.38 \pm 0.07$ & $7.63_{-0.09}^{+0.07}$ & $-0.54_{-0.26}^{+0.25}$ & 2, 3, 9\\ 
Mrk 590& $28.8_{-4.2}^{+3.6}$  & $43.65 \pm 0.06$ & $7.55_{-0.07}^{+0.05}$ & $-0.13_{-0.25}^{+0.24}$ & 2, 3, 9\\  
Mrk 6& $18.5_{-2.4}^{+2.5}$  & $43.71 \pm 0.08$ & $8.03_{-0.06}^{+0.06}$ & $-1.01_{-0.16}^{+0.17}$ & 10, 11\\ 
3C 390.3 & $23.6_{-6.7}^{+6.2}$  & $43.68 \pm 0.10$ & $8.87_{-0.15}^{+0.10}$ & $-3.35_{-0.65}^{+0.60}$ & 2, 3, 12\\ 
3C 390.3 & $46.4_{-3.2}^{+3.6}$  & $44.50 \pm 0.03$ & $9.20_{-0.03}^{+0.03}$ & $-2.12_{-0.51}^{+0.51}$ & 2, 3, 13\\ 
\enddata
\tablecomments{The black hole masses for NGC 2617$^{a}$ and NGC 3516$^{a}$ are adopted from results calculated using $\sigma_{\rm line}$, due to their peculiar line profiles. For the remaining objects, we adopt results derived from the FWHM with a virial factor of $f = 1$. References: (1) \citet{Bentz2006}, (2) \citet{Bentz2013}, (3) \citet{Du2019}, (4) \citet{Feng2021a}, (5) \citet{Denney2010}, (6) \citet{Feng2021b}, (7) \citet{Fausnaugh2017}, (8) \citet{Santos1997}, (9) \citet{Peterson1998}, (10) \citet{Du2018a}, (11) \citet{Yang2024}, (12) \citet{Dietrich1998}, (13) \citet{Dietrich2012}.
}
\end{deluxetable*}

The intercalibration was performed using the PyCALI \citep{Li2014, Pycali2024} program, which employs a Bayesian approach assuming constant AGN component color and host-galaxy starlight during observations to compute flux differences between bands. This assumption is applicable to our data because: (1) the wavelength coverage of the $B$, $V$, and $g$ bands are close, so color variations can be ignored; (2) the Lijiang and ASAS-SN data utilize large fixed apertures for photometry, which can effectively mitigate the variation of host galaxy contribution with seeing; and (3) the ALeRCE data have already corrected the host galaxy contribution during image subtraction. The long-term light curve from ZTF is extracted using PSF photometry, where the contribution of an extended galaxy relative to the point source may vary with seeing. However, given a typical $g$-band seeing of $\sim$2\farcs1 \citep{Bellm2019} and the relatively weak host galaxy of NGC 4151 compared to the AGN \citep[see][and Figure \ref{fig2}]{Chen2023}, the variation in the host galaxy contribution should only fluctuate within a small range in most cases.

For all data, we initially convert all photometric data into flux using the STMAG system for convenience. Then, we select the $g$-band data from ASAS-SN as the reference for the long-term light curve and the $B$-band data from Lijiang as the reference for the RM continuum, and perform intercalibration on other datasets respectively. The RM data are calibrated separately for each year. The long-term light curve is shown in Figure \ref{fig1}, while Figure \ref{fig3} and Table \ref{tab:2} present the RM continuum for each year accordingly.

\subsection{Spectroscopic Measurements} \label{sec:3.2}
The optical spectrum of AGN is a blend of various components, including broad and narrow emission lines, continuum, \feii\ emissions, host-galaxy starlight, and possibly a Paschen continuum. Accurate analysis of the emission-line properties and measurement of the AGN luminosity requires disentangling these different components from the observed spectrum.

Following the method outlined in Paper I, we employ spectral fitting using the DASpec software \citep{Du2024} to decompose the spectra, and the fitting strategy is essentially consistent across the observing seasons. Since Paper I completed the fitting analysis of the data from S1, we focus on the spectra from the subsequent three seasons in this work. In the new fitting process, we allow the power-law index to be a free parameter to better accommodate variations in the continuum shape. This is motivated by the fact that during the new observing period, NGC 4151 entered a bright state and exhibited dramatic variability. Such behavior can lead to non-negligible time lags between the continuum at different wavelengths \citep{Zhou2024} and potential changes in the relative contribution of the Paschen continuum.

Based on the fitted width of the \oiii$\lambda$5007 line in our mean spectrum, and comparing it to the intrinsic width reported by \citet{Whittle1992}, we infer instrumental broadenings of 850 $\kms$ and 1260 $\kms$ for Grism 14 with slit widths of 2\farcs5 and 5\farcs05, respectively. The light curves of the broad \ha, \hb, \hg, \hei, and \heii\ are directly extracted from the fitted broad-line profiles, as shown in Figure \ref{fig3}, and the corresponding data are listed in Table \ref{tab:2}. The AGN monochromatic luminosity at 5100 \AA\ ($L_{\rm 5100}$) and its uncertainty are calculated from the mean and standard deviation of the power-law flux at this wavelength ($\lambda f_{\lambda}$) for each observing season, as detailed in Table \ref{tab:3}.

\begin{deluxetable*}{lcccccccccccc}[!ht]
 \tablecolumns{12}
\tablewidth{\textwidth}
\tabletypesize{\scriptsize}
\tablecaption{Time Lags, Line Widths, and Black hole Masses \label{tab:4}}	
\tablewidth{\textwidth}
\tablehead{& &   \multicolumn{2}{c}{CCF}  && \multicolumn{2}{c}{Mean} &&  \multicolumn{2}{c}{rms}  \\
\cline{3-4} \cline{6-7} \cline{9-10} 
\colhead{Season} & \colhead{Line} & \colhead{$r_{\rm max}$} & \colhead{$\tau_{\rm cent} $} && \colhead{$\rm FWHM$} & \colhead{$\sigma_{\rm line}$} & &  \colhead{$\rm FWHM$} & \colhead{$\sigma_{\rm line}$} & \multicolumn{1}{@{}c@{}}{$M_{\rm VP}^{a}$} &\multicolumn{1}{@{}c@{}}{\mbh}  & \multicolumn{1}{@{}c@{}}{$M_{\rm VP}^{b}$}\\ 
& &  & \multicolumn{1}{c}{(days)}  && \multicolumn{2}{c}{($\kms$)} && \multicolumn{2}{c}{($\kms$)} &  \multicolumn{3}{c}{($\times 10^7 M_{\odot}$)}
   }		
\startdata
S1& \ha & 0.81 &  $5.00_{-3.80}^{+0.84}$&& $4805 \pm 5$ & $2040 \pm 2$ && $4864 \pm 444$ & $2247 \pm 84$ & $0.49_{-0.38}^{+0.09}$ & $3.10_{-2.48}^{+0.93}$ & $2.25_{-1.71}^{+0.38}$ \\
& \hb & 0.88 &  $4.95_{-1.20}^{+2.12}$ && $4937 \pm 14$ & $2095 \pm 6$ && $5950 \pm 636$ & $2623 \pm 54$ & $0.67_{-0.16}^{+0.29}$ & $4.19_{-1.44}^{+2.06}$ & $2.36_{-0.57}^{+1.01}$ \\
& \hg & 0.81 &  $4.90_{-2.61}^{+1.99}$&& $4824 \pm 19$ & $2050 \pm 8$ && $6106 \pm 587$ & $2625 \pm 56$ & $0.66_{-0.35}^{+0.27}$ & $4.15_{-2.43}^{+1.96}$ & $2.22_{-1.18}^{+0.90}$ \\
& \hei & 0.91 &  $1.46_{-1.52}^{+0.86}$&& $5840 \pm 52$ & $2506 \pm 22$ && $8752 \pm 510$ & $3383 \pm 62$ & $0.33_{-0.34}^{+0.19}$ & $2.06_{-2.19}^{+1.31}$ & $0.97_{-1.01}^{+0.57}$ \\
& \heii & 0.90 &  $0.23_{-1.39}^{+1.76}$&& $7106 \pm 55$ & $3001 \pm 21$ && $8900 \pm 352$ & $3631 \pm 80$ & $0.06_{-0.36}^{+0.45}$ & $0.38_{-2.26}^{+2.85}$ & $0.23_{-1.37}^{+1.73}$ \\
\hline
S2& \ha & 0.91 &  $10.27_{-4.91}^{+0.82}$  && $4903 \pm 12$ & $2084 \pm 5$ && $4237 \pm 59$ & $1953 \pm 43$ & $0.76_{-0.37}^{+0.07}$ & $4.82_{-2.58}^{+1.23}$ & $4.82_{-2.31}^{+0.39}$ \\
& \hb & 0.94 &  $9.77_{-3.55}^{+1.40}$  && $5163 \pm 17$ & $2167 \pm 7$ && $4665 \pm 56$ & $2092 \pm 42$ & $0.83_{-0.30}^{+0.12}$ & $5.26_{-2.29}^{+1.48}$ & $5.08_{-1.85}^{+0.73}$ \\
& \hg & 0.92 &  $4.32_{-2.27}^{+2.58}$  && $5233 \pm 28$ & $2232 \pm 12$ && $4813 \pm 110$ & $2330 \pm 64$ & $0.46_{-0.24}^{+0.28}$ & $2.89_{-1.67}^{+1.86}$ & $2.31_{-1.21}^{+1.38}$ \\
& \hei & 0.88 &  $1.68_{-0.51}^{+5.87}$  && $5906 \pm 53$ & $2541 \pm 23$ && $5546 \pm 287$ & $2765 \pm 80$ & $0.25_{-0.08}^{+0.88}$ & $1.58_{-0.61}^{+5.54}$ & $1.14_{-0.35}^{+4.00}$ \\
& \heii & 0.95 &  $4.71_{-3.54}^{+2.83}$  && $6731 \pm 127$ & $3003 \pm 75$ && $9220 \pm 526$ & $3880 \pm 206$ & $1.38_{-1.05}^{+0.84}$ & $8.71_{-6.93}^{+5.71}$ & $4.16_{-3.13}^{+2.51}$ \\
\hline
S3& \ha & 0.84 &  $10.65_{-2.15}^{+1.27}$  && $4671 \pm 5$ & $1981 \pm 2$ && $6265 \pm 435$ & $2445 \pm 48$ & $1.24_{-0.26}^{+0.16}$ & $7.83_{-2.46}^{+2.11}$ & $4.54_{-0.92}^{+0.54}$ \\
& \hb & 0.87 &  $7.32_{-1.12}^{+0.88}$  && $5066 \pm 13$ & $2154 \pm 5$ && $5369 \pm 331$ & $2596 \pm 42$ & $0.96_{-0.15}^{+0.12}$ & $6.06_{-1.73}^{+1.63}$ & $3.67_{-0.56}^{+0.44}$ \\
& \hg & 0.79 &  $6.53_{-1.37}^{+0.96}$  && $5280 \pm 19$ & $2251 \pm 9$ && $7711 \pm 349$ & $2936 \pm 67$ & $1.10_{-0.24}^{+0.17}$ & $6.93_{-2.22}^{+1.96}$ & $3.56_{-0.74}^{+0.52}$ \\
& \hei & 0.89 &  $6.38_{-0.88}^{+0.91}$ && $5384 \pm 30$ & $2316 \pm 16$ && $4279 \pm 126$ & $2911 \pm 59$ & $1.05_{-0.15}^{+0.16}$ & $6.64_{-1.85}^{+1.86}$ & $3.61_{-0.50}^{+0.52}$ \\
& \heii & 0.85 &  $0.49_{-0.51}^{+0.52}$ && $7753 \pm 46$ & $3315 \pm 19$ && $9268 \pm 340$ & $3917 \pm 61$ & $0.15_{-0.15}^{+0.16}$ & $0.92_{-0.98}^{+1.01}$ & $0.57_{-0.60}^{+0.61}$ \\
\hline
S4& \ha & 0.68 &  $3.77_{-1.27}^{+1.37}$  && $5137 \pm 4$ & $2180 \pm 2$ & &$7195 \pm 277$ & $2679 \pm 42$ & $0.53_{-0.18}^{+0.19}$ & $3.33_{-1.38}^{+1.45}$ & $1.94_{-0.66}^{+0.70}$ \\
& \hb & 0.78 &  $3.25_{-0.72}^{+1.40}$  && $5407 \pm 7$ & $2297 \pm 3$ & &$6146 \pm 536$ & $2765 \pm 43$ & $0.48_{-0.11}^{+0.21}$ & $3.05_{-1.00}^{+1.51}$ & $1.85_{-0.41}^{+0.80}$ \\
& \hg & 0.68 &  $2.35_{-1.70}^{+2.07}$  && $6263 \pm 26$ & $2662 \pm 11$ & &$9248 \pm 492$ & $3559 \pm 77$ & $0.58_{-0.42}^{+0.51}$ & $3.66_{-2.80}^{+3.34}$ & $1.80_{-1.30}^{+1.58}$ \\
& \hei & 0.24 &  $4.56_{-1.51}^{+0.98}$  && $6171 \pm 19$ & $2624 \pm 8$ & &$6743 \pm 397$ & $3121 \pm 55$ & $0.87_{-0.29}^{+0.19}$ & $5.46_{-2.24}^{+1.76}$ & $3.39_{-1.12}^{+0.72}$ \\
& \heii & 0.72 &  $0.45_{-0.95}^{+0.78}$  && $8931 \pm 79$ & $3848 \pm 32$ & &$10944 \pm 898$ & $4810 \pm 94$ & $0.20_{-0.43}^{+0.35}$ & $1.27_{-2.71}^{+2.23}$ & $0.70_{-1.47}^{+1.21}$ \\
\enddata
\tablecomments{The time lags are given in the rest frame, and line widths are corrected for instrumental broadening. $M_{\rm VP}^{a}$ and $M_{\rm VP}^{b}$ represent the virial products calculated using $\sigma_{\rm line}$ and FWHM, respectively. \mbh\ denotes the black hole mass derived from $\sigma_{\rm line}$, using the virial factor from \citet{Ho2014}. The uncertainties in \mbh\ incorporate contributions from the time lag, line width, and the virial factor $f$.
}
\end{deluxetable*}

\subsection{Lags, SMBH Masses, and Accretion Rates} \label{sec:3.3}
The time lag between the continuum and emission-line light curves is measured using the interpolated cross-correlation function \citep[ICCF;][]{Gaskell1987}. In calculating the correlation coefficients, we interpolate the emission-line and continuum light curves separately and take the average of their calculated results as the final ICCF. The time delay is determined by the centroid ($\tau_{\rm cent}$) of the region above 80\% of the ICCF peak ($r_{\rm max}$). To estimate the uncertainty in the time lag, we employ a combined Monte Carlo ``flux randomization" and bootstrap ``random subset selection" method, generating a cross-correlation centroid distribution \citep[CCCD;][]{Peterson2004} with 10,000 realizations. The 15.87\% and 84.13\% positions of the CCCD serve as the lower and upper limits of the 1$\sigma$ error. The ICCF and CCCD results are shown in Figure \ref{fig3}, and the values of $r_{\rm max}$ and $\tau_{\rm cent}$ are listed in Table \ref{tab:4}. It is clear that the BLR of NGC 4151 exhibits a stratified radial structure each year.

The characteristic velocity of the BLR is represented by the full width at half maximum (FWHM) and velocity dispersion ($\sigma_{\rm line}$) of the root mean square (rms) spectrum, corrected for instrumental broadening. Uncertainties are obtained via the bootstrap method (see Paper I for details). Considering the low spectral resolution and the sensitivity of rms to outliers, we directly use the fitted broad emission line profiles to calculate the rms spectrum. Using Equation \ref{equ1}, we derive the black hole mass for each year. Note that the $v$ is not equivalent to $\sigma_{\rm line}$ or FWHM, and the calculated value is referred to as the virial product black hole mass ($M_{\rm VP}$). The final black hole mass needs to be corrected by multiplying a virial factor $f$, which accounts for the geometry and kinematics of the BLR. In this work, we use $f = 6.3\pm1.3$ \citep{Ho2014} to obtain the \mbh\ based on $\sigma_{\rm line}$. For FWHM-based calculations, we use $f = 1$, consistent with \citet{Du2019}, to facilitate a systematic analysis of the properties of CL-AGNs.

The dimensionless accretion rate of the AGN is calculated using the following formula\citep{Du2018b}:
\begin{equation}\label{equ:accre}
\dotm = \frac{\dot{\mbh} c^{2}}{L_{\rm Edd}} = 20.1\,\left(\frac{L_{44}}{\cos i}\right)^{3/2}m_7^{-2},
\end{equation}
where $\dot{\mbh}$ denotes the mass accretion rates of the black hole, $L_{\rm Edd}$ is the Eddington luminosity, $L_{44}=L_{5100}/10^{44}\,\ergs$, $m_7 = \mbh/10^7\,\sunm$, and $i$ is the inclination angle of the accretion disk. 
The inclination angle of NGC 4151, as derived from its NLR, is $\sim 45^\circ$ \citep[cos $i = 0.71$;][]{Das2005, Fischer2013}, which agrees well with the average inclination angle of $\sim 40^\circ$ \citep[cos $i = 0.77$;][]{Nandra2007} found in AGNs using Fe K$\alpha$ measurements. Given that the inclination angle is not measured for most AGNs, and to facilitate comparison with the results in \citet{Du2018b}, we adopt the same average value of cos $ i = 0.75$. The results show that during the four years of observations, the $\dotm$ of NGC 4151 ranges from 0.01 to 0.23. Table \ref{tab:4} presents the broad emission-line widths and black hole masses, while the accretion rates are given in Table \ref{tab:3}.

\begin{figure*}[!ht]
\centering 
\includegraphics[scale=0.6]{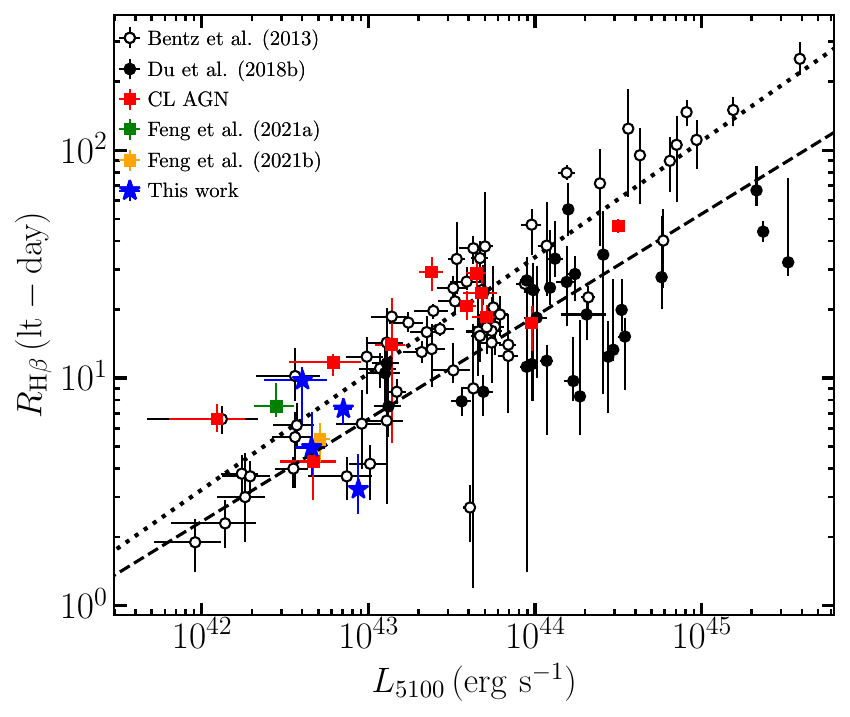}
\includegraphics[scale=0.6]{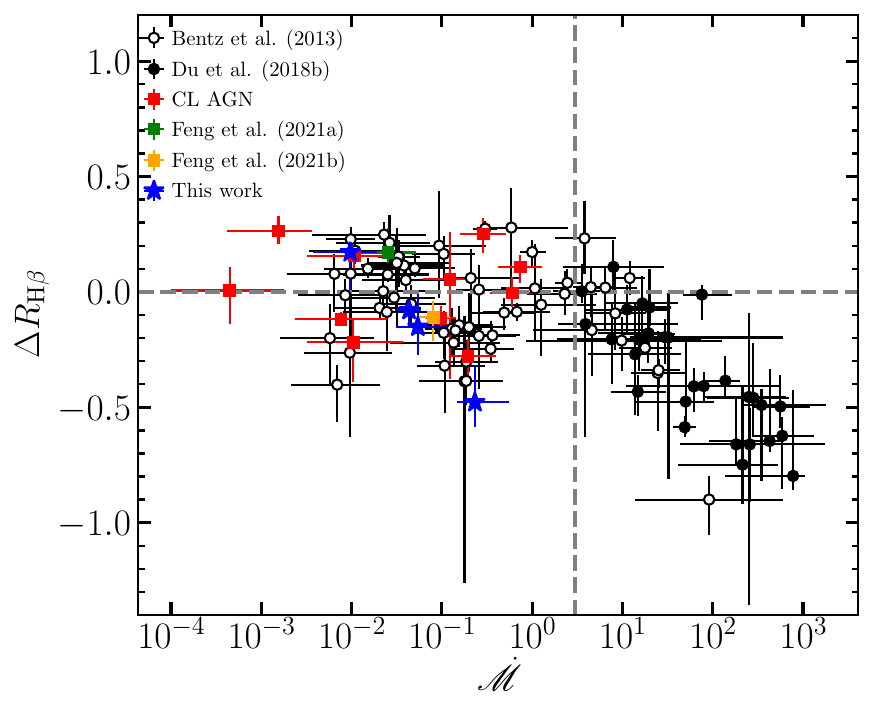}
\caption{Correlations of $R_{\rm H\beta}$ vs. $L_{\rm 5100}$ (left) and $\Delta R_{\rm H\beta}$ vs. $\dotm$ (right). To ensure a consistent comparison, all luminosities have been recalculated following \citet{Du2019}. In the left panel, the dotted and dashed lines represent the linear fit from \citet{Du2018b} for AGNs with $\dotm < 3$ and $\dotm \geqslant 3$, respectively. In the right panel, the horizontal and vertical grey dashed lines indicate $\Delta R_{\rm H\beta} = 0$ and $\dotm = 3$, respectively.
}
\label{fig4}
\end{figure*}

\subsection{Examination of $R-L$ Relation in CL-AGN} \label{sec:3.4}
Examining the change in the BLR size of AGN with luminosity is crucial for understanding the $R-L$ relation. We compared the \hb\ radius ($R_{\rm H\beta}$) and $L_{\rm 5100}$ for each observing season with the $R-L$ relation established for objects with $\dotm < 3$ by \citet{Du2018b}, which closely resembles the relation reported by \citet{Bentz2013}. The results indicate that during our observations, NGC 4151 exhibited an ``anti-breathing" phenomenon \citep{Gilbert2003, Goad2004, Cackett2006, Wang2020}, where the time delay of the broad emission line decreased as the luminosity increased. The data points from the four observing seasons exhibit considerable scatter, though they do not deviate from the locations of AGNs with similar luminosity.

\citet{Du2018b} found that when $\dotm \geqslant$ 3, the BLR lag can be shorter than the expected value from the $R-L$ relation, and this offset of lag is related to the accretion rate. Accordingly, we also examined the relationship between the relative offset of the \hb\ radius ($\Delta R_{\rm H\beta} = {\rm log}(R_{\rm H\beta}/ R_{{\rm H\beta},R-L})$) and \dotm. We found that NGC 4151 is consistent with AGNs where $\dotm < 3$, suggesting that the CL phenomenon may be associated with lower accretion rates. To verify that this is not a unique property of NGC 4151, we compiled an RM sample of CL-AGNs (Table \ref{tab:3}) and conducted the same analysis. As shown in Figure \ref{fig4}, all CL-AGNs are sub-Eddington accretors, and their distribution on the $R_{\rm H\beta}$ vs. $L_{\rm 5100}$ plot is similar to that of AGNs with $\dotm < 3$. This is also consistent with the fact that most CL-AGNs are classified as Population B sources in the Eigenvector 1 schema, as Population B is generally associated with lower accretion rates \citep{Panda2024}.

\begin{figure*}[!ht]
\centering 
\includegraphics[scale=0.32]{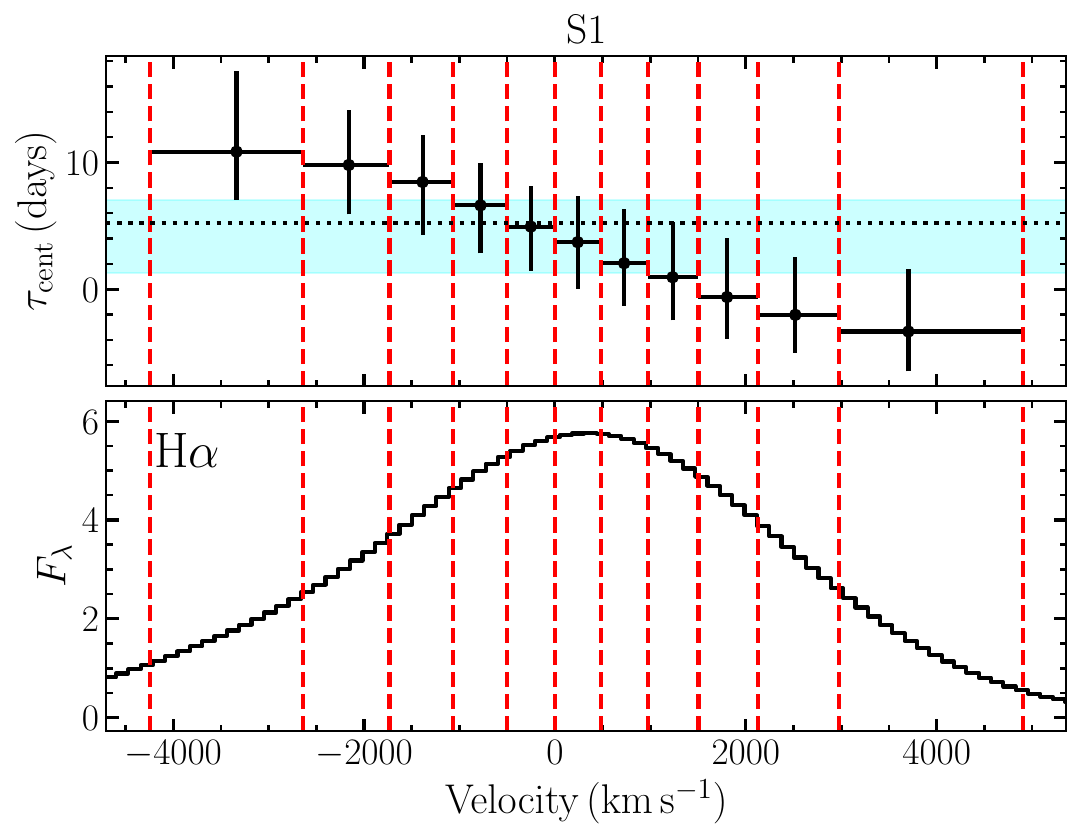}
\includegraphics[scale=0.32]{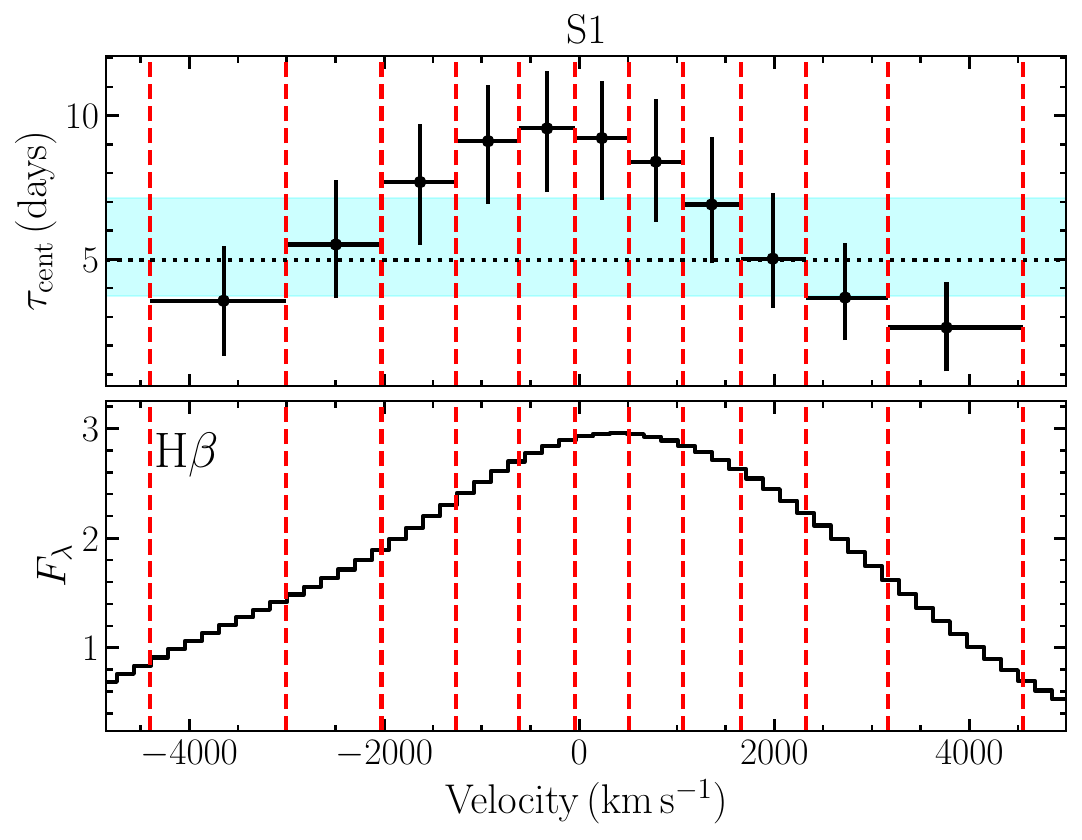}
\includegraphics[scale=0.32]{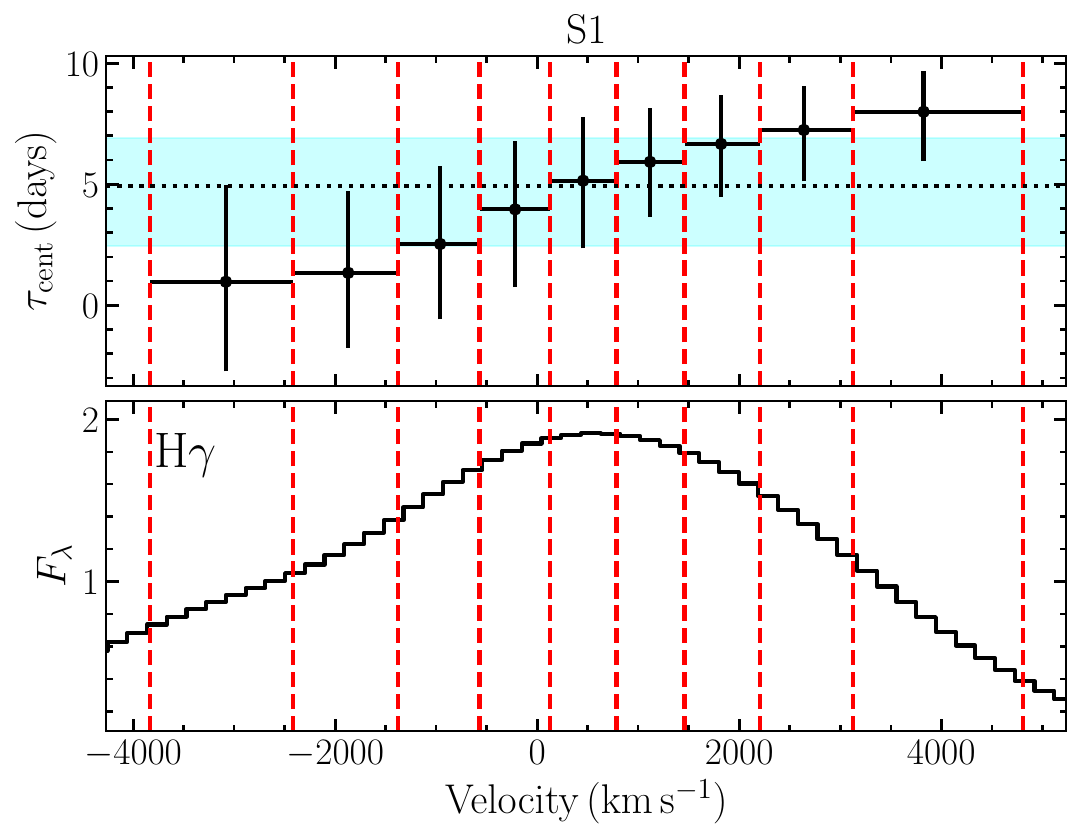}
\includegraphics[scale=0.32]{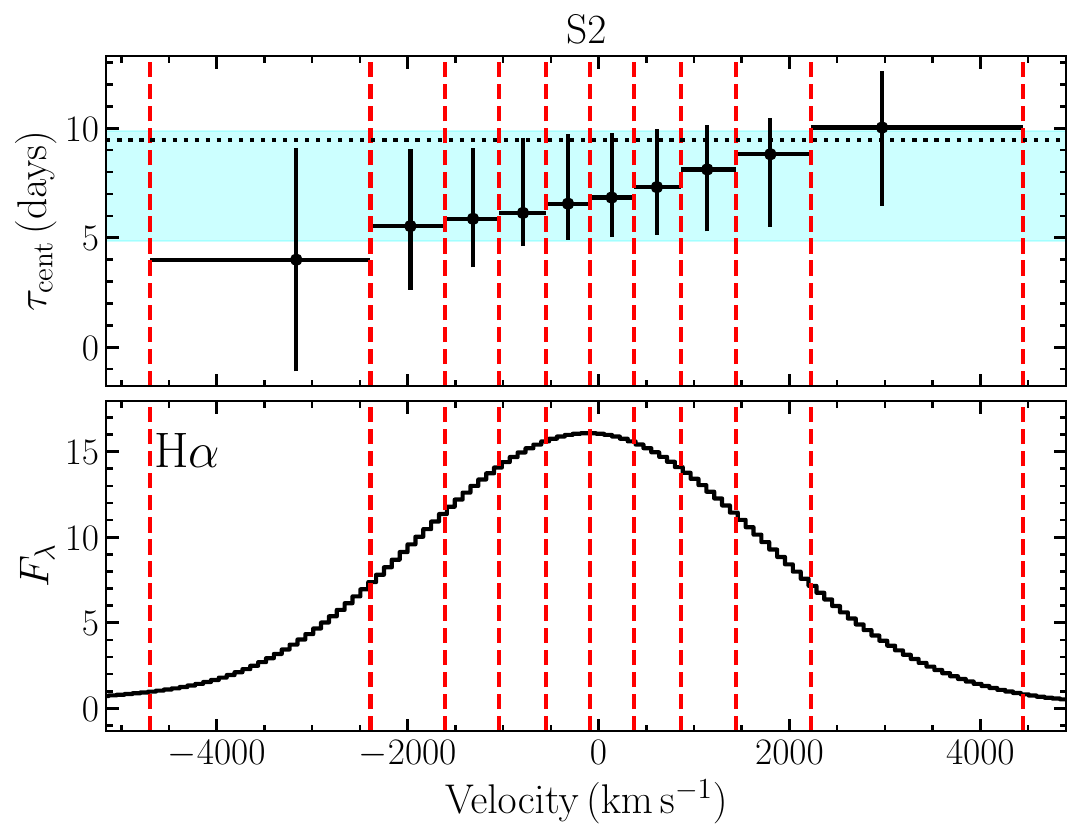}
\includegraphics[scale=0.32]{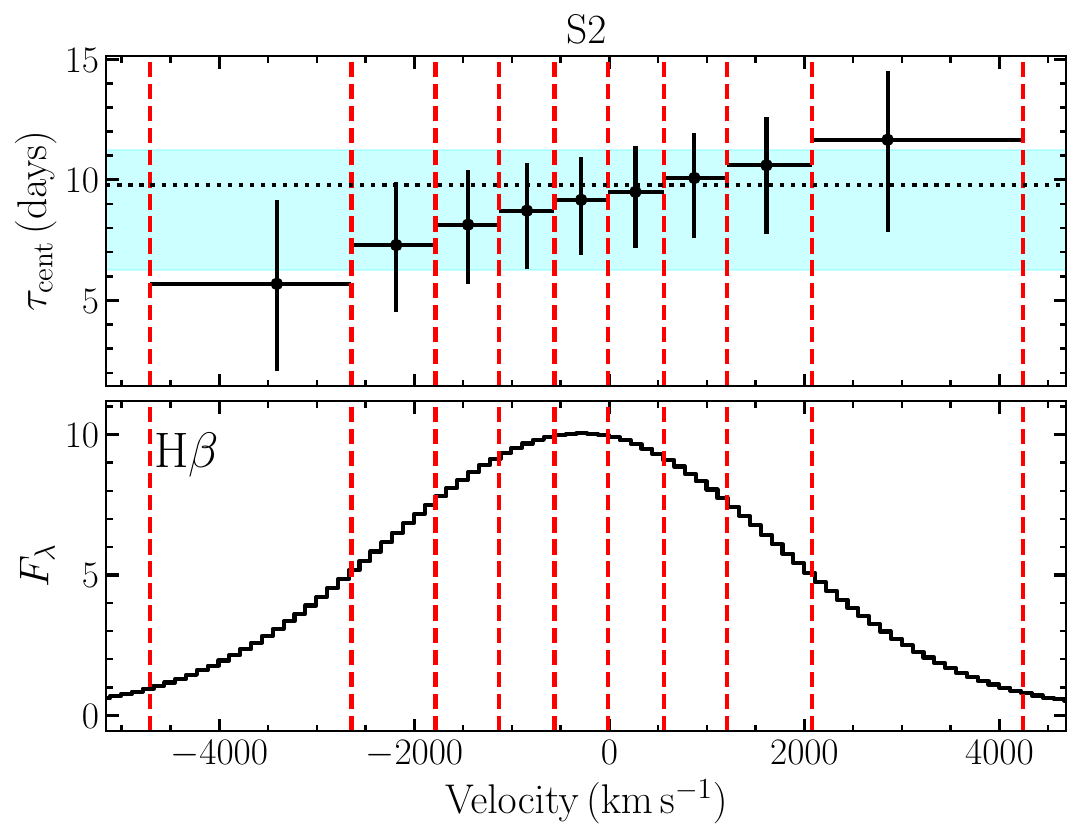}
\includegraphics[scale=0.32]{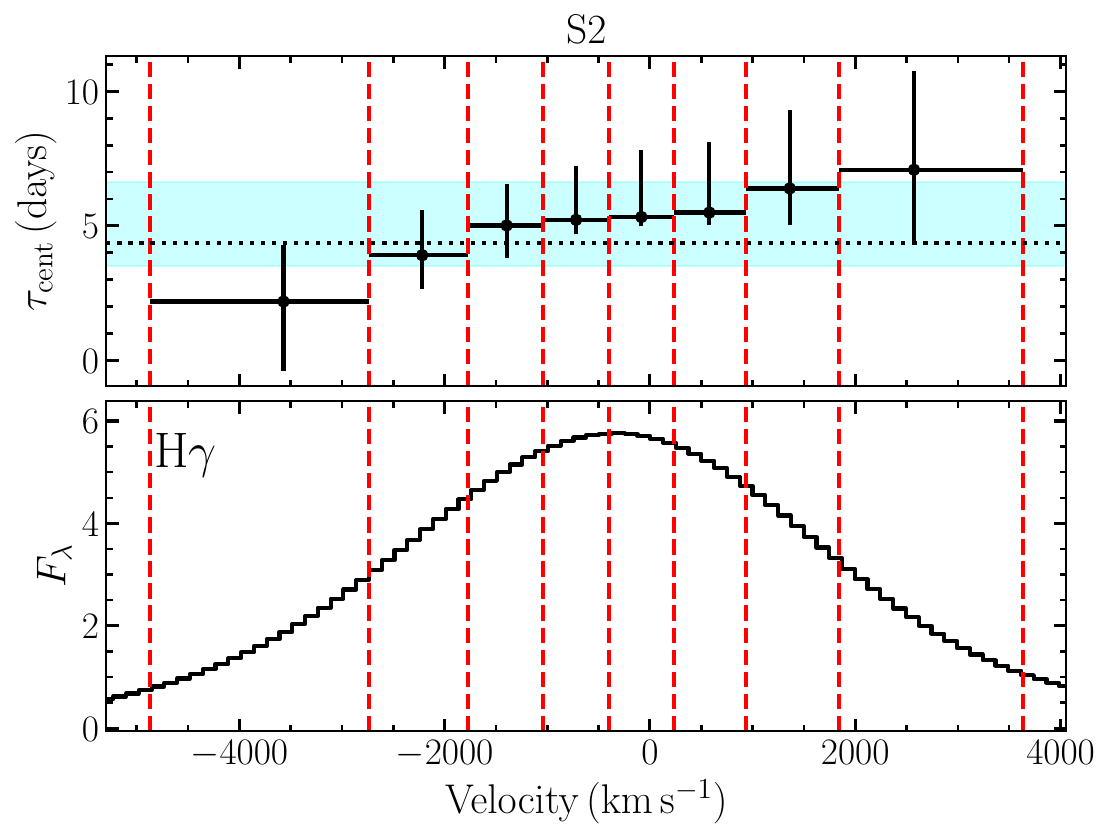}
\includegraphics[scale=0.32]{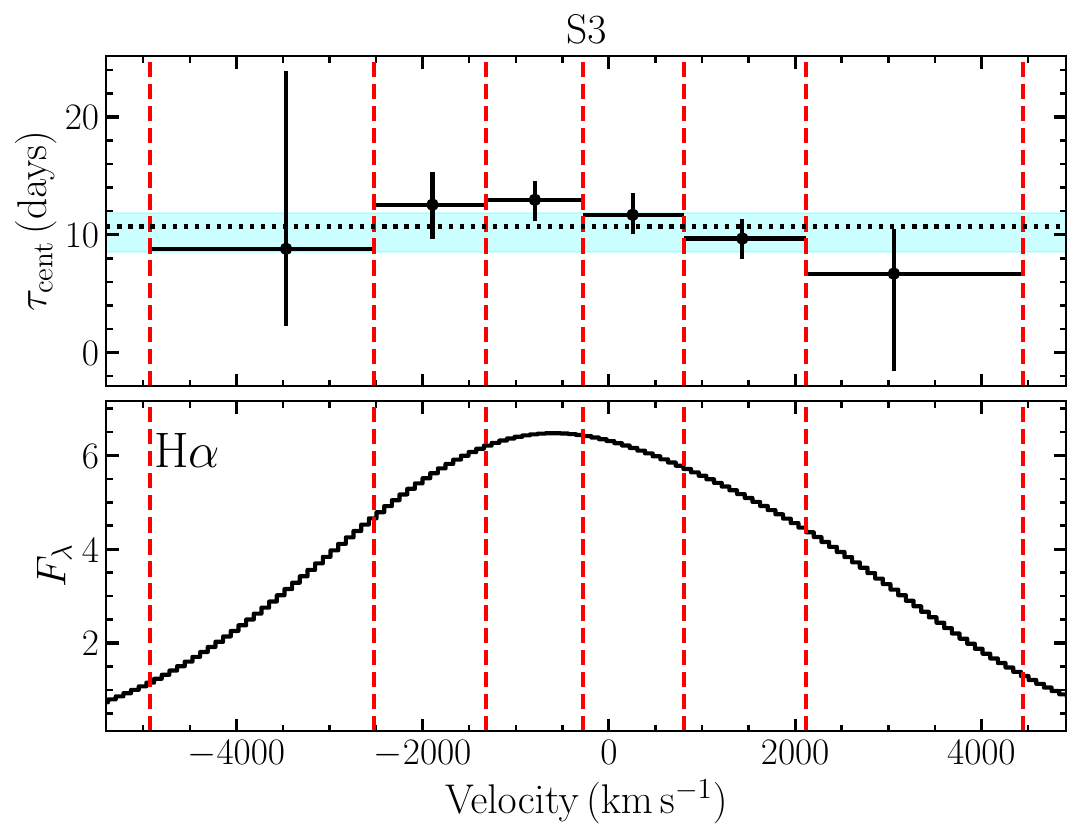}
\includegraphics[scale=0.32]{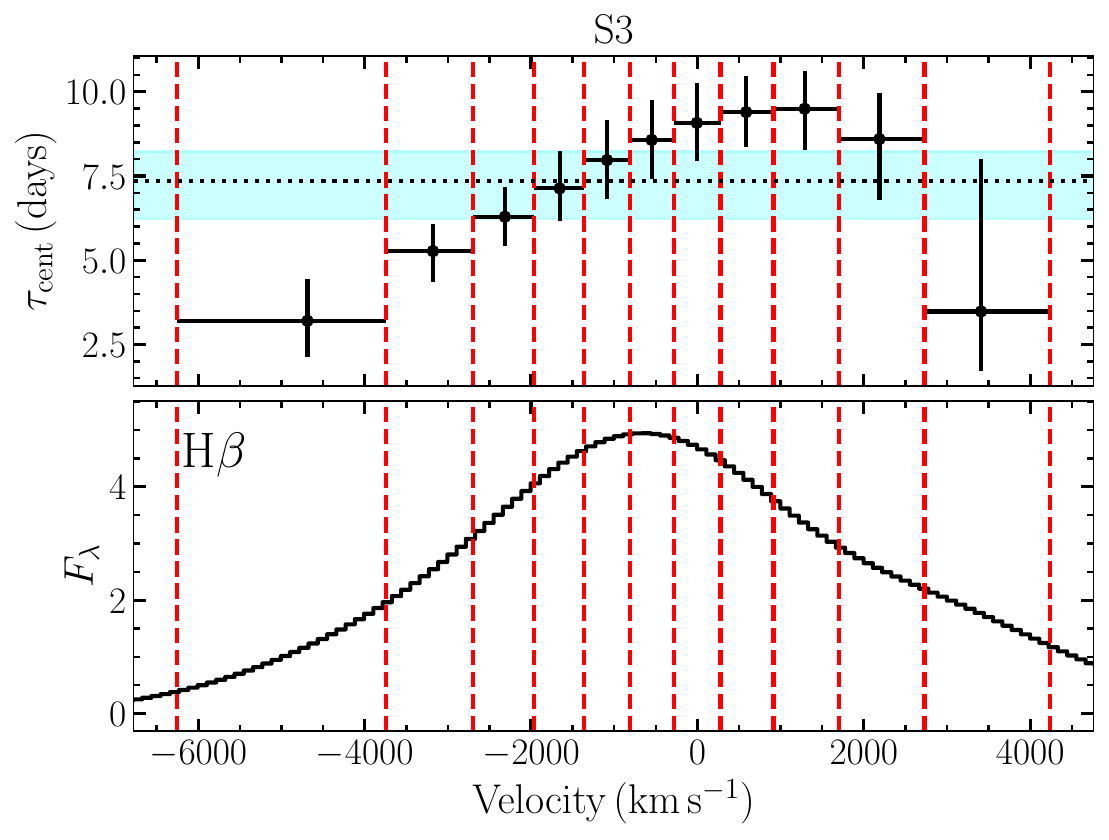}
\includegraphics[scale=0.32]{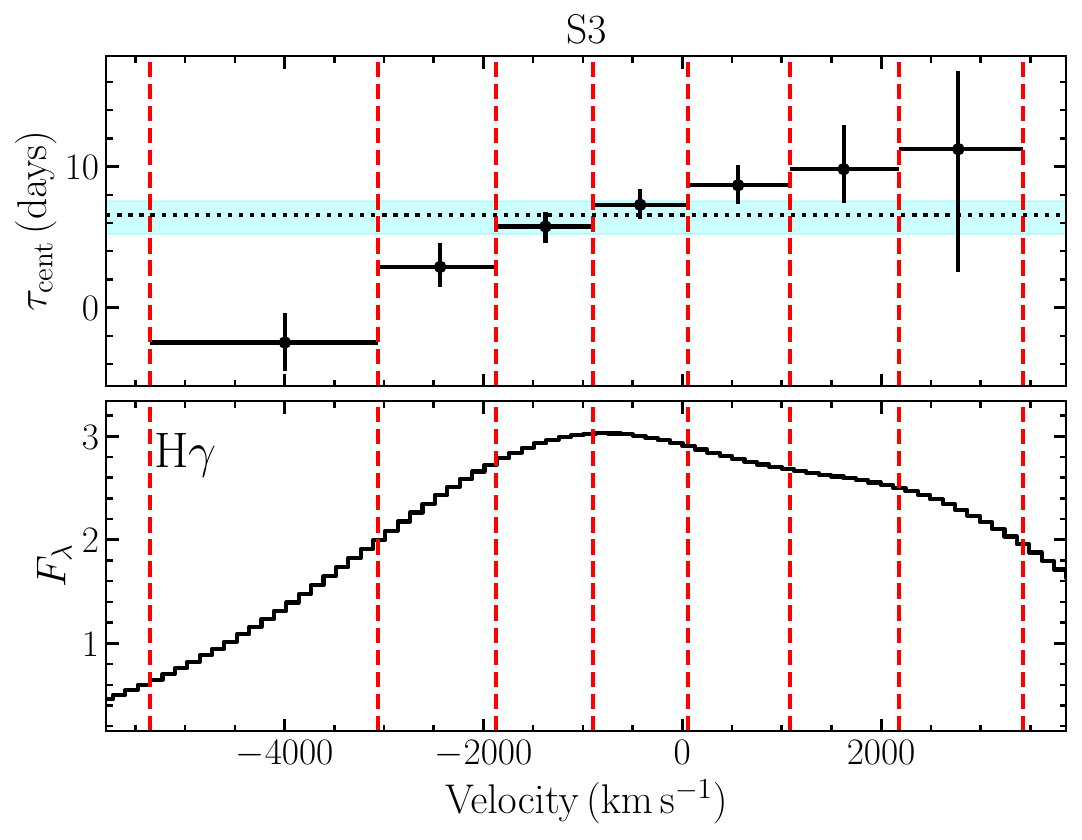}
\includegraphics[scale=0.32]{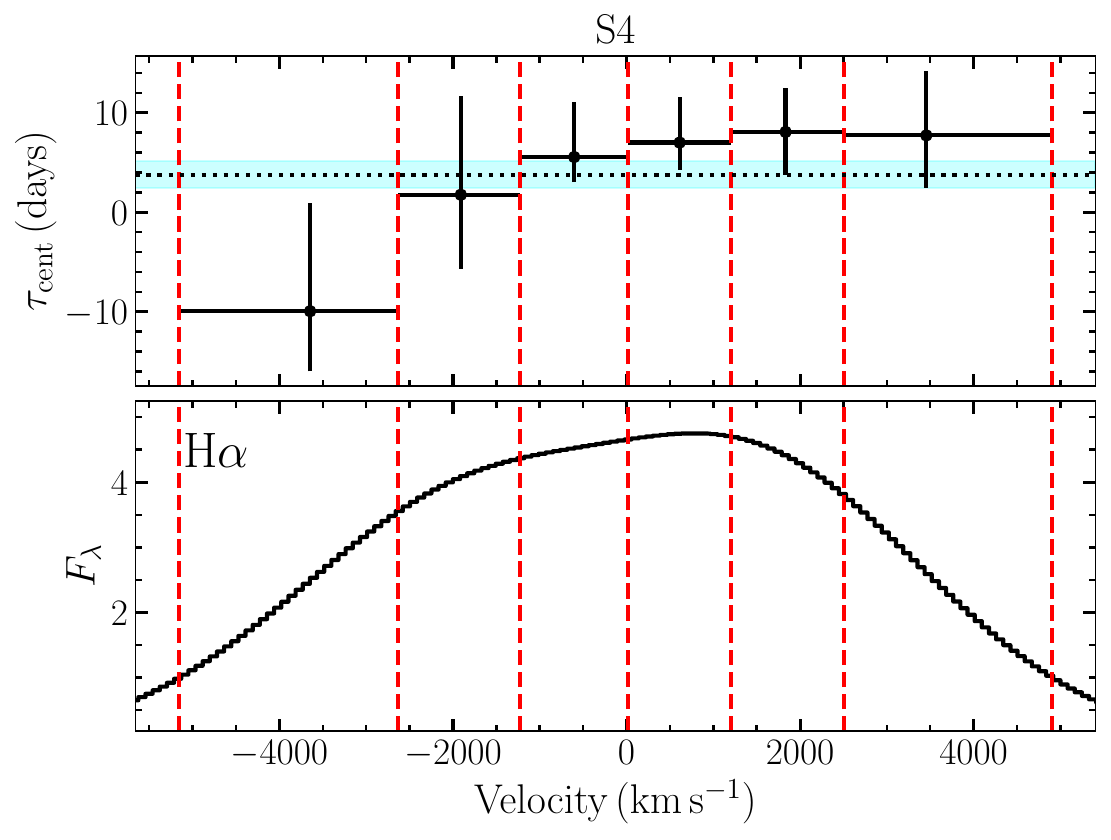}
\includegraphics[scale=0.32]{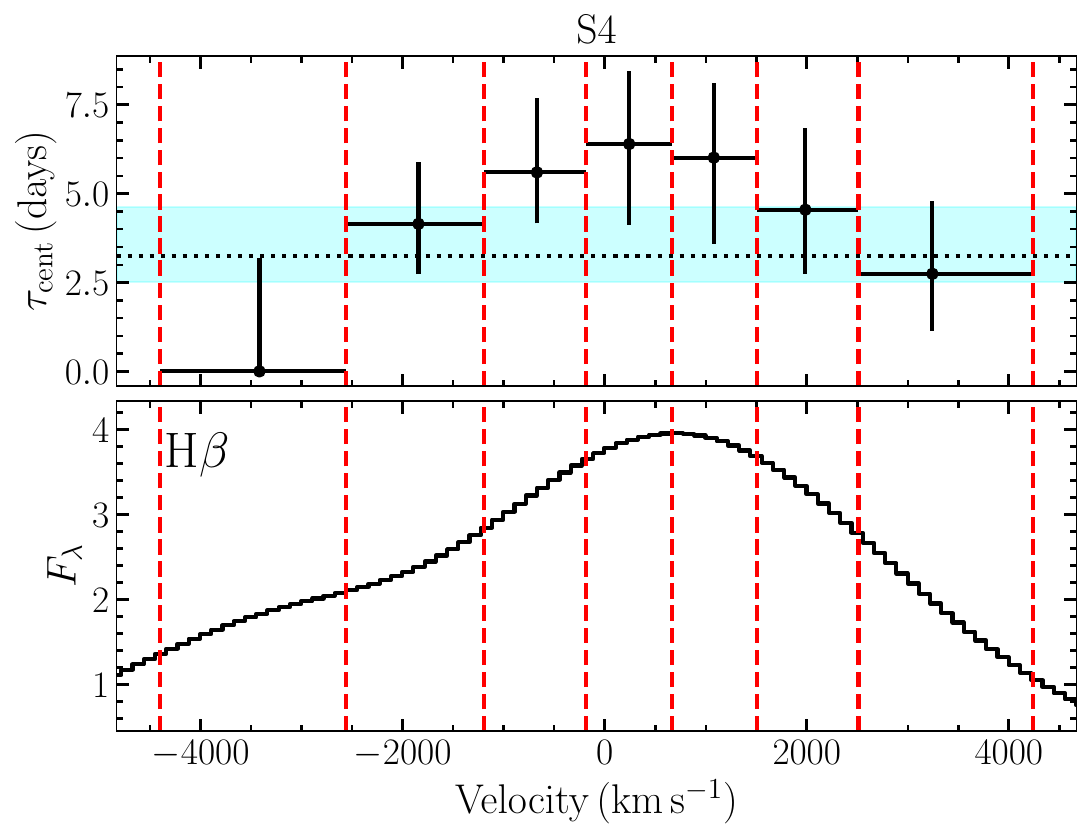}
\includegraphics[scale=0.32]{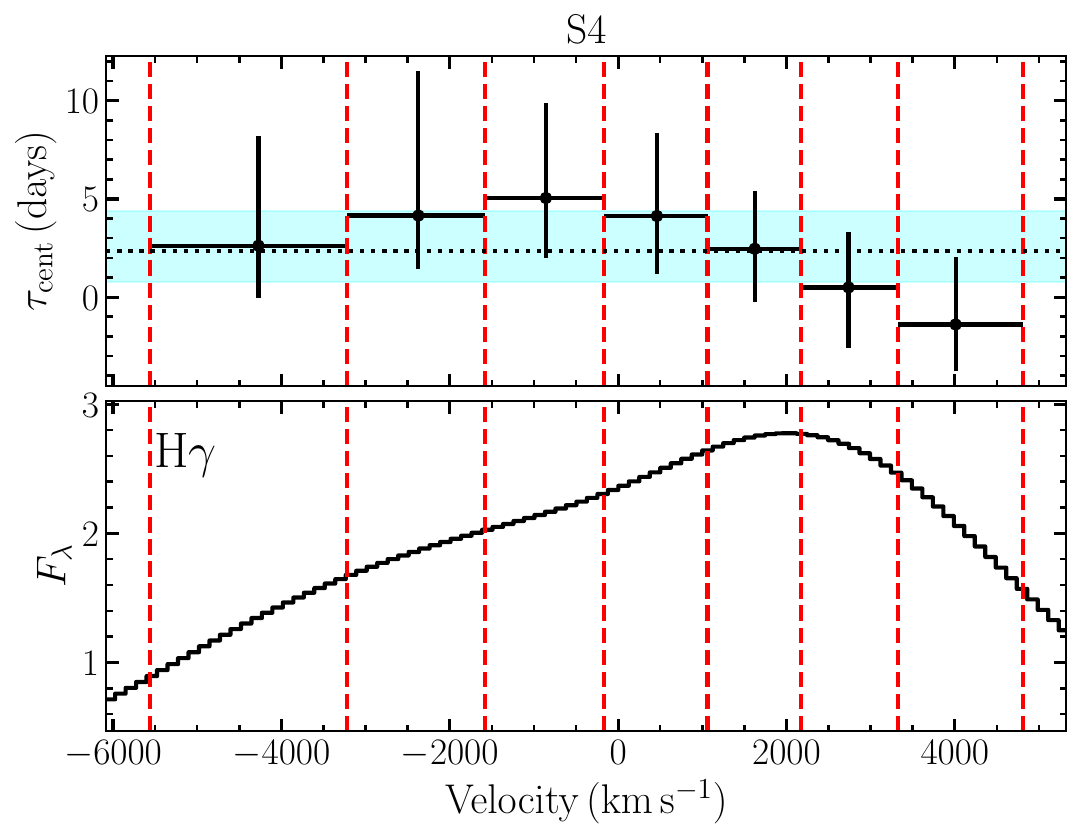}
\caption{Velocity-resolved lags for \ha, \hb, and \hg\ in each observing season. In each panel, the lower plot displays the rms spectrum, while the upper plot shows the observed $\tau_{\rm cent}$ for each flux bin, divided by the vertical red dashed lines. The horizontal dotted line and the associated green-shaded region represent the overall lag of the emission line and its corresponding uncertainty.
}
\label{fig5}
\end{figure*}

\begin{figure*}[!ht]
\centering 
\includegraphics[scale=0.415]{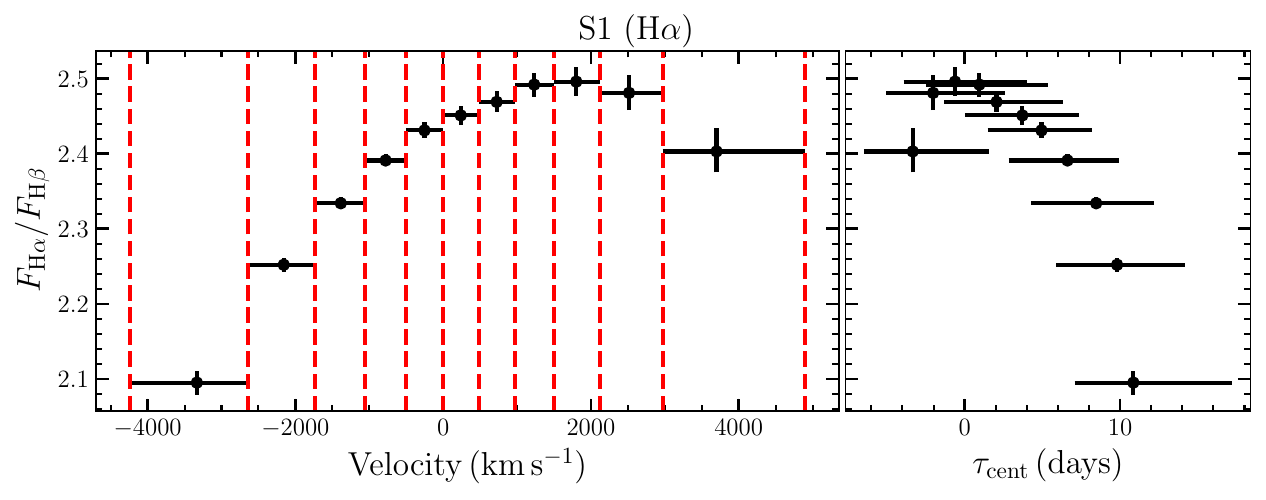}
\includegraphics[scale=0.415]{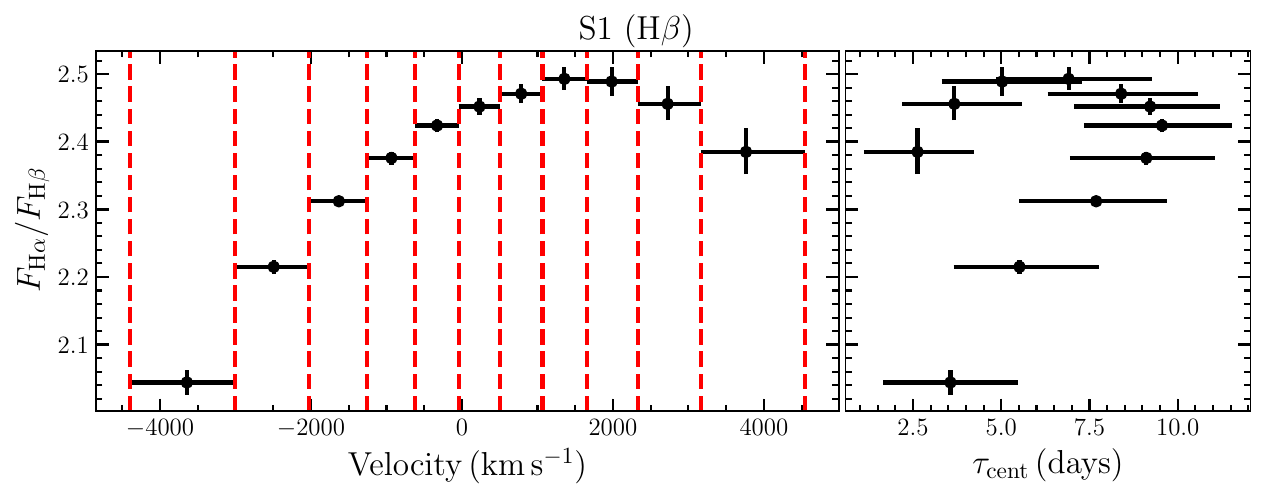}
\includegraphics[scale=0.415]{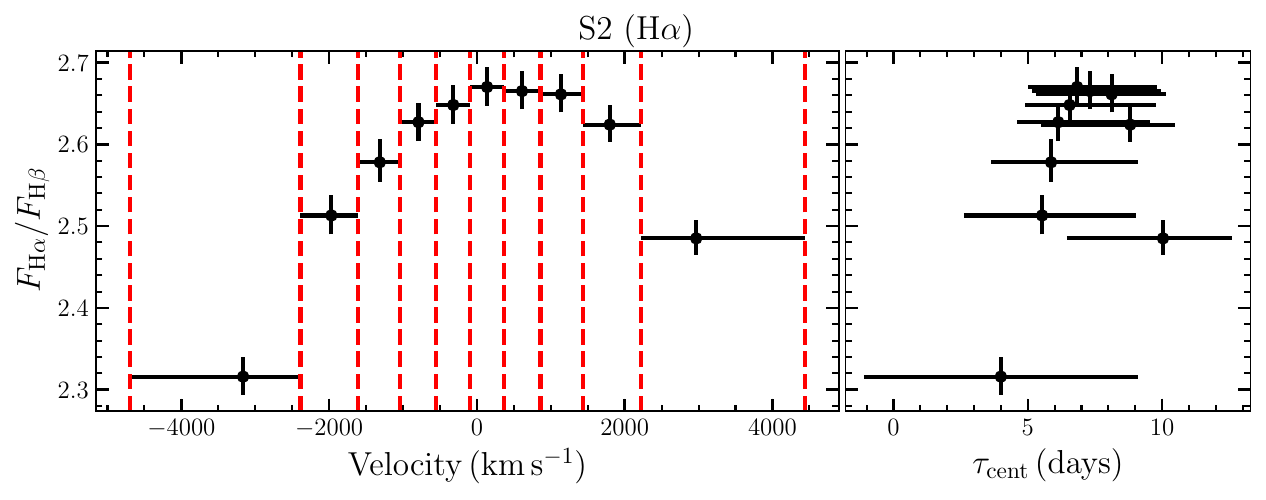}
\includegraphics[scale=0.415]{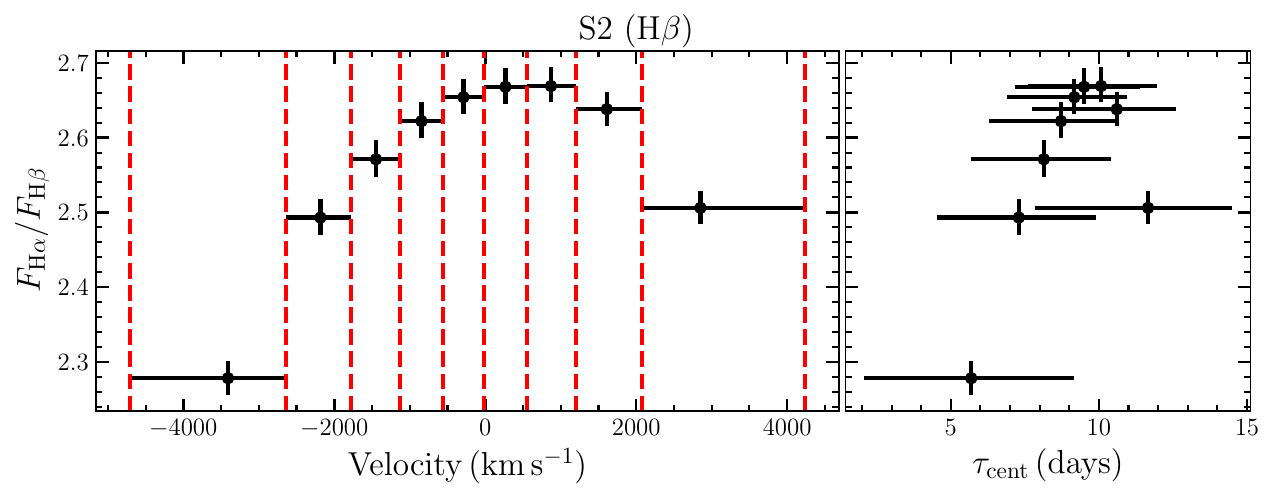}
\includegraphics[scale=0.415]{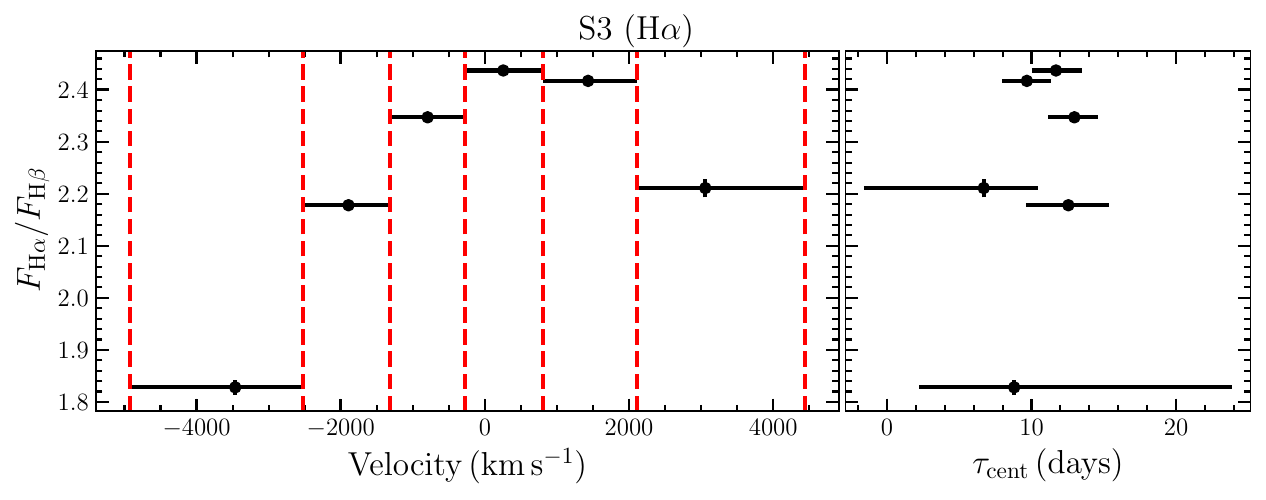}
\includegraphics[scale=0.415]{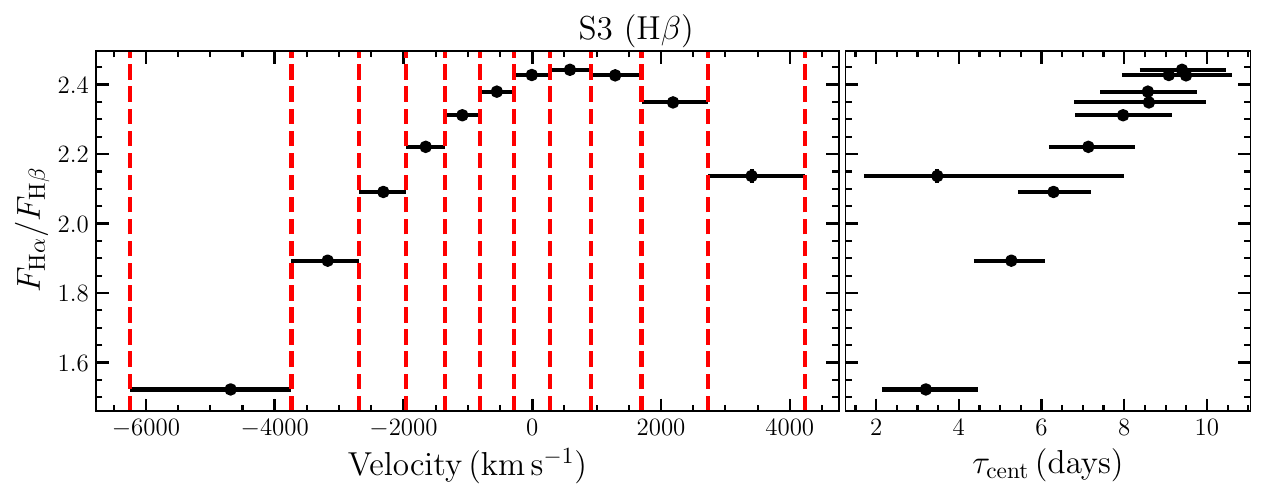}
\includegraphics[scale=0.415]{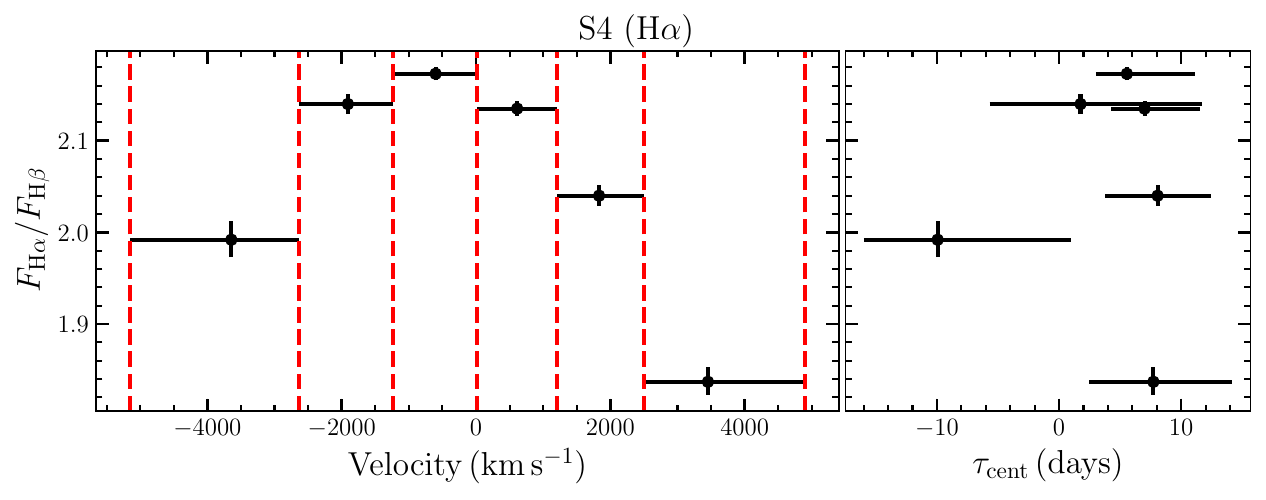}
\includegraphics[scale=0.415]{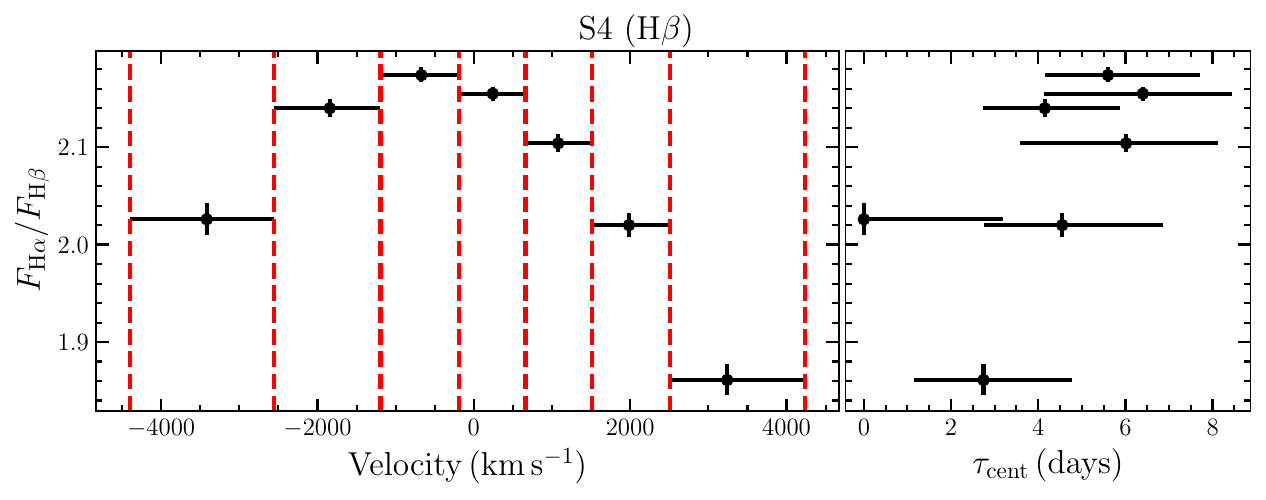}
\caption{Velocity-resolved BDs in each observing season. The panels in the left column present the results calculated using the velocity bins of \ha, while those in the right column show the results using the velocity bins of \hb. In each panel, the plot on the right side illustrates the correlation between the lags and BDs for the corresponding emission line across different velocity bins.
}
\label{fig6}
\end{figure*}

\subsection{Velocity-resolved Signatures} \label{sec:3.5}
We employed two methods—velocity-resolved RM and velocity-resolved IM—to analyze the distribution and velocity field of BLR gas in NGC 4151 for each observing season. Velocity-resolved RM captures variations in photon travel paths by measuring the time delays across different velocities within the BLR relative to the ionization source \citep{Peterson1993}. On the other hand, velocity-resolved IM traces the distance of gas at varying velocities from the center by measuring the flux ratios of emission lines \citep{Li2024}. Combining these two methods can effectively mitigate the degeneracy between BLR geometry and kinematics.

Since measuring velocity-resolved lags requires a high S/N of the spectra, we only analyzed the strongest emission lines: \ha, \hb, and \hg. For the line ratios, we selected the commonly used \fab, also known as the Balmer decrement (BD). The measurements of velocity-resolved lags and BDs were achieved by the fitted broad emission line profiles. To ensure similar flux variations within each velocity bin, we divided the rms spectra of the emission lines into several bins based on equal flux. The number of bins was chosen to maintain a balance between the S/N of the light curve and the preservation of as much velocity information as possible. As a result, the number of velocity bins varies for different emission lines across observing seasons. It should be noted that the calculation of line ratios requires consistent velocity bins for the two emission lines. Therefore, we performed separate calculations based on the rms profiles of \ha\ and \hb. The BD values and uncertainties are obtained by the average and random sampling as described in \citet{Li2024}. 

Figure \ref{fig5} shows the measurements of velocity-resolved lags for \ha, \hb, and \hg\ across each observing season. All three emission lines exhibit variability, with \ha\ and \hb\ showing more rapid changes. In contrast, \hg\ only demonstrates a different velocity-resolved lag pattern in the final year compared to the previous three years. Figure \ref{fig6} presents the results of velocity-resolved BDs, which also exhibit changes across different observing seasons. These results are evident that the geometry and kinematics of BLR in NGC 4151 are undergoing rapid changes on timescales of months to years.

\begin{figure}[!ht]
\centering 
\includegraphics[scale=0.45]{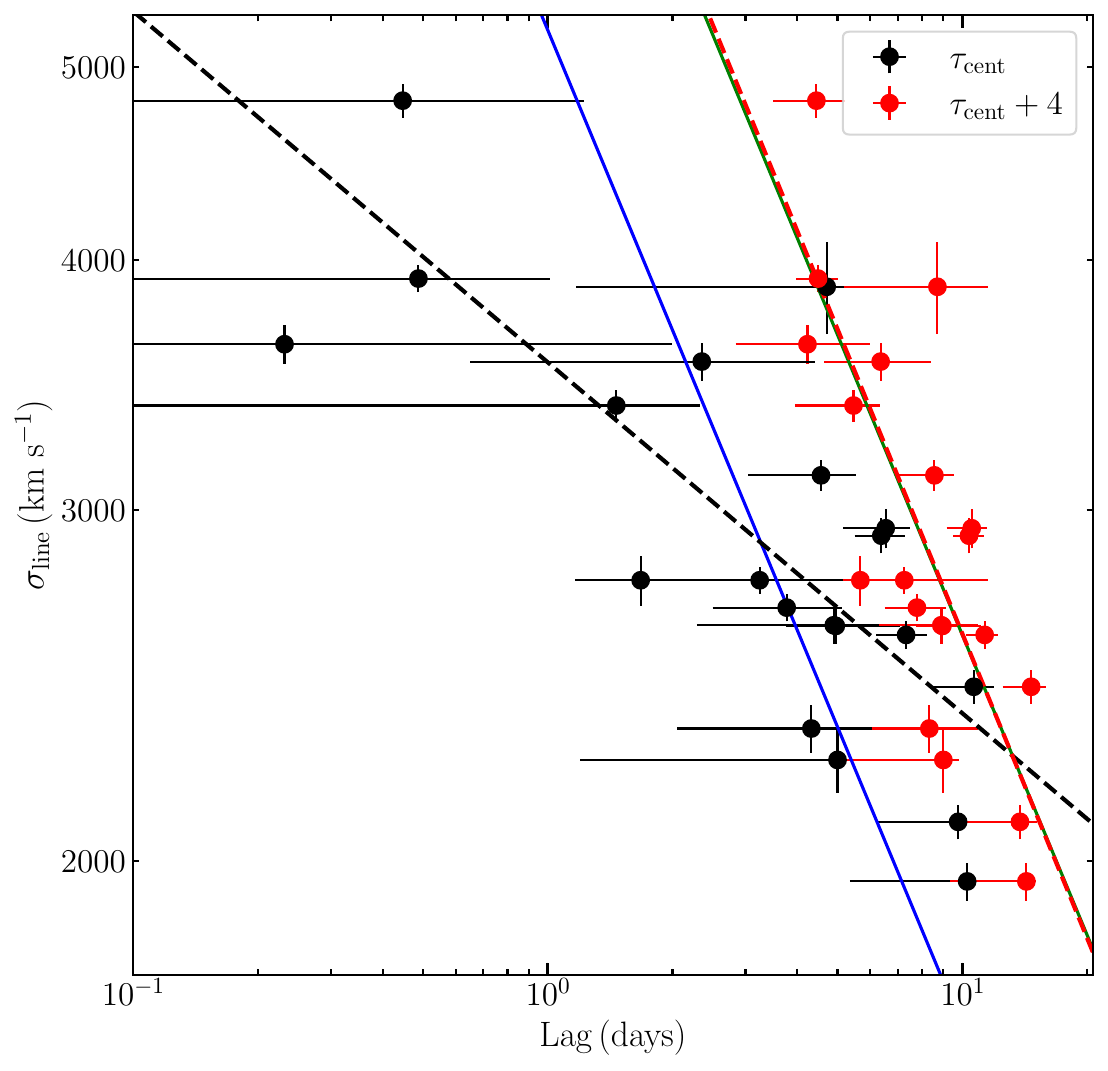}
\caption{The relationship between $\sigma_{\rm line}$ and lags for all emission lines in our observations. The black points represent the observed $\tau_{\rm cent}$ of the emission lines listed in Table \ref{tab:4}, while the red points include an additional 4-day correction to account for the continuum lag between optical and UV. The black and red dashed lines represent fits to the black and red points, respectively, using the relation $\sigma_{\rm line} \propto \tau_{\rm cent}^{\beta}$. The blue and green solid lines show the fits with $\beta$ fixed at 0.5.
}
\label{fig7}
\end{figure}

\section{Discussion} \label{sec:4}
\subsection{Potential Impact of UV-Optical Continuum Lags} \label{sec:4.1}
Photoionization simulations predict that the time delays of the Balmer and Helium emission lines in the BLR will ``breathe" with changes in luminosity, and different emission lines will exhibit radial stratification due to the effects of optical depth and ionization energy \citep{Korista2004}. These phenomena have been observed in some AGNs and used to test the virial relationship of the BLR \citep[e.g.,][]{Kollatschny2014, Fausnaugh2017, Wang2020, Chen2023}. In all observation seasons of NGC 4151, the lags of \ha, \hb, \hg, \hei, and \heii\ almost always show a trend of $\tau_{\rm H\alpha} > \tau_{\rm H\beta} > \tau_{\rm H\gamma} > \tau_{\rm{He}\,\textsc{i}} > \tau_{\rm{He}\,\textsc{ii}}$, which is consistent with theoretical expectations. In contrast, the lags of the broad emission lines relative to luminosity exhibit a completely opposite ``breathing" phenomenon (see Table \ref{tab:3} and Figure \ref{fig4}). This anomalous ionization behavior may be linked to peculiar motions within the BLR. Indeed, our analysis reveals significant discrepancies in \mbh\ measurements across different observing seasons, and the black hole masses derived from different emission lines also vary. Furthermore, the velocity-resolved signatures deviate from the predictions of the virial motion. These phenomena suggest the possible presence of a non-virialized BLR in NGC 4151.

To determine the dominant motion of the BLR gas, we compared the time delays and line widths of all the emission lines (Figure \ref{fig7}). The best-fit result ($v \propto \tau ^{-0.18\pm 0.05}$) significantly deviates from the virial relationship ($v \propto \tau ^{-0.5}$), which is inconsistent with our previous findings. In Paper I, by collecting historical observations, we found that the BLR of NGC 4151 is essentially virialized. However, that study accounted for the time delay between optical and ultraviolet (UV) continuum emissions, and recently, \citet{Zhou2024} found that the continuum lag is comparable to the BLR lag in bright states. This effect becomes more pronounced for emission lines closer to the accretion disk, such as Helium and higher-order Balmer lines. To account for this, we arbitrarily added 4 days to the lags of each emission line, based on the continuum lag of 4.78$\pm$0.63 days at 5100 \AA\ for S3 calculated by \citet{Zhou2024}. The corrected fitting result ($v \propto \tau ^{-0.51\pm 0.04}$) agrees well with the virial relationship. This leads us to favor the interpretation that the anomalous ``breathing" phenomenon and the inconsistent \mbh\ measurements may be caused by the influence of the UV-optical lag. A more comprehensive analysis, including simultaneous measurements of the continuum and BLR lags, will be investigated in our future work.

\subsection{Evolution of BLR Geometry and Kinematics} \label{sec:4.2}
NGC 4151 exhibited significantly different velocity-resolved signatures across four observing seasons, suggesting that the geometry and/or kinematics of emitting gas clouds, or the ionizing radiation field, may be changing. These variations can generally be explained by the following possibilities: (1) the presence of peculiar substructures in the BLR \citep[e.g., spiral arm or hot spot,][]{Storchi-Bergmann2017}, whose distribution changes as the gas moves; (2) the acceleration/deceleration processes driven by variations in radiation pressure associated with changes in the accretion rate \citep{Netzer2010, Czerny2011}; and (3) the evolution of an anisotropic ionizing source \citep[e.g., a binary SMBH system,][]{Ji2021}.

Changes in the BLR structure are often linked to dynamical timescales ($\tau_{\rm dyn}$) or orbital timescales ($\tau_{\rm orb}$). The $\tau_{\rm dyn}$ corresponds to the time needed for significant radial motion of the gas, while $\tau_{\rm orb}$ represents the period required for the gas to complete one orbit around the central source. Assuming that the gas in the BLR is entirely in radial or circular motion (without requiring Keplerian motion), we can directly estimate the minimum $\tau_{\rm dyn}$ and $\tau_{\rm orb}$ using the lag and width of the emission line, respectively:
\begin{equation}
\tau_{\rm dyn} = \frac{c \tau}{v} = \frac{1.6\, R_{\rm 10}}{v_{\rm 5000}}\,\rm{yr},
\end{equation}
\begin{equation}
\tau_{\rm orb} = \frac{2 \pi c \tau}{v} = \frac{10.3\, R_{\rm 10}}{ v_{\rm 5000}}\,\rm{yr},
\end{equation}
where $R_{10}$ is the BLR radius in units of 10 lt-days, and $v_{\rm 5000}$ is the emission line width in units of 5000 $\kms$. For a BLR with a radius of 10 lt-days (considering the UV-optical lag) and a velocity of 5000 $\kms$ (FWHM), we obtain $\tau_{\rm dyn} \approx$ 1.6 years and $\tau_{\rm orb} \approx$ 10.3 years. The orbital timescale is significantly longer than the observed changes, especially for the velocity-resolved lags of \ha. The dynamical timescale appears to be more consistent with observations, and the velocity-resolved BDs also support the hypothesis that the structure of the emitting region is changing.

However, there are still two issues that are challenging to resolve. First, the actual $\tau_{\rm dyn}$ should be much longer than our calculated value, since our calculation assumes that the gas is entirely in radial motion, which is unlikely in the virial-dominated BLR discussed in Section \ref{sec:4.1}. Second, the timescale of changes in the inner BLR should be shorter, the emitting clouds of higher-order Balmer lines are closer to the central SMBH, with higher velocities and smaller radius, and should yield shorter $\tau_{\rm dyn}$ values. Yet, as shown in Figure \ref{fig5}, the velocity-resolved lag results suggest a timescale order of \ha $\leqslant$ \hb $\leqslant$ \hg, contrary to expectations.

The large amplitude of long-term variability shown in Figure \ref{fig1} provides a possible explanation: increasing radiation pressure effectively accelerates the radial motion of BLR gas, thereby changing its kinematic properties. In this framework, emission lines with longer time delays should be less affected. This is because that both radiation pressure and gravity decrease with increasing distance from the center, following the $1/R_{\rm BLR}^{2}$, leading to a similar decrease in acceleration. Such expectation is contrary to our observations—the emission lines with longer time delays undergo more significant changes. Moreover, as radiation pressure increases, the BLR kinematics should evolve from inflow to outflow. In Figure \ref{fig5}, assuming that the changes in velocity-resolved lags are only related to kinematics, \hb\ and \hg\ even exhibit the opposite trend to what is expected, evolving from outflow to virial motion as increasing radiation pressure.

If there is a change in the illuminating direction of the central ionizing source, the variability timescales of different emission lines should differ by only a few days, corresponding to their time delays. Meanwhile, the velocity-resolved lags and BDs should vary similarly and correlate with luminosity. For a binary black hole system, we would also expect periodic variations. Currently, these phenomena have not been detected in our data.

The aforementioned contradictions suggest that the rapid changes in the velocity-resolved signatures of NGC 4151 may be attributed to multiple mechanisms. This implies that the structure, kinematics, and ionizing field of the BLR may be more complex than expected, necessitating further observational evidence to uncover the underlying physical processes.

\subsection{Implication of CL-AGN} \label{sec:4.3}
The physical origin of the CL phenomenon may be attributed to significant changes in the obscuring material along the line of sight or intrinsic changes in the accretion rate of the central engine, both of which can cause dramatic variations in AGN luminosity \citep{Ricci2023}. These two mechanisms can be distinguished by observable quantities that are sensitive or insensitive to obscuration, such as variations in infrared and hard X-ray flux \citep{Sheng2017, Temple2023}, or changes in X-ray absorption column density \citep{Matt2003}. The RM method can also impose additional constraints to further probe the physical properties of CL-AGNs.

If the CL phenomenon is caused by changes in the obscuring material, the size of the BLR should remain relatively constant, despite substantial variations in the AGN luminosity. In this case, the lag of the CL-AGN would lie above the normal AGN $R-L$ relation. As depicted in Figure \ref{fig4}, the spectroscopically confirmed CL-AGNs agree with the $R-L$ relation established by low accretion rate AGNs \citep{Du2018b}. This consistency suggests that changes in the accretion rate are more likely responsible for the observed CL phenomenon in these AGNs. Our finding also supports the notion that AGNs with lower accretion rates are more prone to exhibit larger amplitude variability \citep{Sniegowska2020}.

\section{Conclusion} \label{sec:5}
In this study, we conducted a four-year velocity-resolved RM campaign on the CL-AGN NGC 4151 during its outburst phase. Our goal was to investigate the structure, geometry, and kinematics of its BLR and to explore the physical origins of the CL phenomenon. The key findings are summarized as follows:

\begin{enumerate}
\item Through the time lag measurements of \ha, \hb, \hg, \hei, and \heii\ emission lines, we confirmed the stratified structure of the BLR in NGC 4151, with different emission lines originating from different radial distances. This finding is consistent with photoionization models.

\item Across the four observation seasons, we observed that the lags of the broad emission lines relative to the continuum decreased as the luminosity increased—an ``anti-breathing" phenomenon. This behavior deviates from the typical ``breathing" mode expected in AGNs, where the BLR radius expands with increasing luminosity. This anomaly could be linked to non-virialized motions within the BLR, or the influence of the UV-optical lag. We favor the latter interpretation due to a good virial relationship derived after correcting of the UV-optical lag.

\item The velocity-resolved RM and IM analyses revealed that the geometry and kinematics of the BLR in NGC 4151 may undergo rapid and significant changes over the course of a year. These changes cannot be fully explained by any single mechanism, including an inhomogeneous BLR, variations in radiation pressure, or changes in the illuminated ionizing field. This indicates that the physical properties of the BLR might be more complex than previously thought, likely driven by multiple mechanisms.

\item The \hb\ time lags of NGC 4151 and other spectroscopically confirmed CL-AGNs are in agreement with the $R-L$ relation for AGNs with low accretion rates, suggesting that the CL phenomenon is more likely driven by intrinsic changes in the accretion rate rather than by obscuration along the line of sight.

\end{enumerate}

In summary, our study sheds light on the complex internal processes of NGC 4151, providing valuable observational evidence for understanding the origins of the CL phenomenon in AGNs. Long-term, multi-line velocity-resolved RM has proven to be a powerful tool for probing the evolution of the structure and kinematics of the BLR. In future work, we plan to construct two-dimensional velocity-resolved maps and further analyze the time lags associated with both the continuum and the BLR in NGC 4151. These efforts will help us better understand the interaction between the accretion disk and the BLR and its impact on the dynamics of the BLR.

\vspace{5mm}
This work is supported by National Key R\&D Program of China (No. 2021YFA1600404), the National Natural Science Foundation of China (grants No. 12303022, 12203096, 12373018, 11991051, 12322303, 12203041, and 12073068), Yunnan Fundamental Research Projects (grants NO. 202301AT070358 and 202301AT070339), Yunnan Postdoctoral Research Foundation Funding Project, Special Research Assistant Funding Project of Chinese Academy of Sciences, the Natural Science Foundation of Fujian Province of China (No. 2022J06002), and the science research grants from the China Manned Space Project with No. CMS-CSST-2021-A06.

We acknowledge the support of the staff of the Lijiang 2.4 m telescope. Funding for the telescope has been provided by Chinese Academy of Sciences and the People’s Government of Yunnan Province.

Based on observations obtained with the Samuel Oschin Telescope 48-inch and the 60-inch Telescope at the Palomar Observatory as part of the Zwicky Transient Facility project. ZTF is supported by the National Science Foundation under Grants No. AST-1440341 and AST-2034437 and a collaboration including current partners Caltech, IPAC, the Oskar Klein Center at Stockholm University, the University of Maryland, University of California, Berkeley , the University of Wisconsin at Milwaukee, University of Warwick, Ruhr University, Cornell University, Northwestern University and Drexel University. Operations are conducted by COO, IPAC, and UW.

This research has made use of the NASA/IPAC Extragalactic Database (NED), which is operated by the Jet Propulsion Laboratory, California Institute of Technology, under contract with NASA.

\vspace{5mm}

\facility{YAO:2.4m}
\software{PyRAF \citep{Pyraf2012}, DASpec \citep{Du2024}, PyCALI \citep{Pycali2024}.}


\end{document}